%% file: bare_jrnl_compsoc.tex
\documentclass[10pt,journal,compsoc,twoside]{IEEEtran}
\usepackage{booktabs} 

\usepackage[numbers,sort&compress]{natbib}
\usepackage{subfigure}
\usepackage{amssymb}
\usepackage{amsmath,bm}
\usepackage{xcolor}
\usepackage{algorithm}
\usepackage{algorithmic}
\usepackage{mdwlist}  
\usepackage{multirow}
\usepackage{ragged2e}

\usepackage{url}
\PassOptionsToPackage{hyphens}{url}
\usepackage[breaklinks,colorlinks,bookmarks=false]{hyperref}
\hypersetup{citecolor=blue,linkcolor=blue,urlcolor=black}

\newtheorem{definition}{Definition}
\newtheorem{insight}{Insight}

\newcommand{\tabincell}[2]{\begin{tabular}{@{}#1@{}}#2\end{tabular}}
\newcommand{\ignore}[1]{}

\ifCLASSINFOpdf
   \usepackage[pdftex]{graphicx}
   \graphicspath{{../pdf/}{../jpeg/}}
   \DeclareGraphicsExtensions{.pdf,.jpeg,.png}
\else
   \graphicspath{{../eps/}}
   \DeclareGraphicsExtensions{.eps}
\fi
\begin{document}
%
\title{VulDeeLocator: A Deep Learning-based Fine-grained Vulnerability Detector}
%
%
%
%

\author{Zhen~Li, 
        Deqing~Zou, 
        Shouhuai~Xu,
        Zhaoxuan~Chen,
        Yawei~Zhu,
        and Hai~Jin, ~\IEEEmembership{Fellow,~IEEE}
\thanks{Corresponding author: Deqing Zou.}
\IEEEcompsocitemizethanks{\IEEEcompsocthanksitem Z. Li is with the National Engineering Research Center for Big Data Technology and System, Services Computing Technology and System Lab, Cluster and Grid Computing Lab, Big Data Security Engineering Research Center, School of Cyber Science and Engineering, Huazhong University of Science and Technology, Wuhan 430074, China, and also with School of Cyber Security and Computer, Hebei University, Baoding, 071002, China.
E-mail: lizhenhbu@gmail.com
\IEEEcompsocthanksitem D. Zou is with the National Engineering Research Center for Big Data Technology and System, Services Computing Technology and System Lab, Cluster and Grid Computing Lab, Big Data Security Engineering Research Center, School of Cyber Science and Engineering, Huazhong University of Science and Technology, Wuhan 430074, China. 
E-mail: deqingzou@hust.edu.cn
\IEEEcompsocthanksitem S. Xu is with the Department of Computer Science, University of Colorado Colorado Springs, Colorado, USA 80918. This work was partly done at
University of Texas at San Antonio. 
E-mail: sxu@uccs.edu.
\IEEEcompsocthanksitem Z. Chen, Y. Zhu, and H. Jin are with the National Engineering Research Center for Big Data Technology and System, Services Computing Technology and System Lab, Cluster and Grid Computing Lab, Big Data Security Engineering Research Center, School of Computer Science and Technology, Huazhong University of Science and Technology, Wuhan 430074, China.
E-mail: \{zhaoxaunchen, yokisir, hjin\}@hust.edu.cn}
}

%
%

\markboth{IEEE TRANSACTIONS ON DEPENDABLE AND SECURE COMPUTING}
{LI \MakeLowercase{\textit{et al.}}: V\MakeLowercase{ul}D\MakeLowercase{ee}L\MakeLowercase{ocator}: A Deep Learning-based Fine-grained Vulnerability Detector}
%



\IEEEtitleabstractindextext{%
\begin{abstract}
\justifying
Automatically detecting software vulnerabilities is an important problem that has attracted much attention from the academic research community.
However, existing vulnerability detectors still cannot achieve the vulnerability detection capability and the locating precision that would warrant their adoption for real-world use. 
In this paper, we present a vulnerability detector 
that can simultaneously achieve a high detection capability and a high locating precision, dubbed \underline{Vul}nerability \underline{Dee}p learning-based \underline{Locator} (VulDeeLocator). 
In the course of designing VulDeeLocator, we encounter 
difficulties including how to accommodate semantic relations between the definitions of types as well as macros and their uses across files, how to accommodate accurate control flows and variable define-use relations, and how to achieve high locating precision. 
We solve these difficulties
by using two innovative ideas: (i) leveraging intermediate code to accommodate extra semantic information,  and
(ii) using the notion of {\em granularity refinement} to pin down locations of vulnerabilities.
When applied to 200 files randomly selected from three real-world software products, VulDeeLocator detects 18 confirmed vulnerabilities (i.e., true-positives). Among them, 16 vulnerabilities correspond to known vulnerabilities; the other two are not reported in the National Vulnerability Database (NVD) but have been ``silently'' patched by the vendor of Libav when releasing newer versions.
\end{abstract}

}

\maketitle

\IEEEdisplaynontitleabstractindextext

%
\IEEEpeerreviewmaketitle

\section{Introduction}
\IEEEPARstart{S}{oftware} vulnerabilities are a major cause of cyber attacks. Unfortunately, vulnerabilities are prevalent
as evidenced by the steady increase of vulnerabilities reported by the {\em Common Vulnerabilities and Exposures} (CVE) \cite{CVE}.
One important approach towards eliminating vulnerabilities is to design {\em vulnerability detectors} to detect (and patch) them.
An ideal vulnerability detector should simultaneously achieve a high {\em detection capability}
and a high {\em locating precision} (i.e., precisely pinning down the vulnerable lines of code).

A popular family of vulnerability detectors is based on {\em static analysis}. These detectors can be divided into {\em code similarity-based} ones and {\em pattern-based} ones.
Code similarity-based detectors \cite{kim2017vuddy,jang2012redebug,li2016vulpecker} can detect vulnerabilities caused by code cloning, and can achieve a high locating precision when they indeed detect vulnerabilities.
However, they incur high false-negatives (i.e., low detection capability) when applied to detect vulnerabilities that are not caused by code cloning.
Pattern-based detectors can be further divided into {\em rule-based} ones and {\em machine learning-based} ones.
Rule-based detectors \cite{FlawFinder,Checkmarx,HP_Fortify,Coverity, DBLP:journals/ijndc/LiangWWX16,DBLP:journals/chinaf/FangLZWWW17}
can identify the vulnerable lines of code when they indeed {\em correctly} detect vulnerabilities,
but often incur a {\em low} detection capability (because of their high false-positives and high false-negatives).
Moreover, they require human analysts to define vulnerability detection rules.
Machine learning-based detectors use vulnerability patterns for detection, where the patterns are learned from analyst-defined feature representation of vulnerable programs
 \cite{yamaguchi2012generalized,neuhaus2007predicting,grieco2016toward,yamaguchi2013chucky,yamaguchi2015automatic}.
However, these detectors cannot achieve a high locating precision because they operate at a coarse granularity,
typically at the function level \cite{yamaguchi2012generalized}.

The recent development in {\em machine learning-based} vulnerability detection is to use deep learning  \cite{vuldeepecker,SySeVR,DBLP:journals/corr/abs-2001-02334}, while operating at a fine-grained {\em program slice} level.
These detectors can relieve the problem of {\em manual-feature definition}, which has received further attention recently \cite{DBLP:conf/ccs/LinZLPX17,DBLP:journals/corr/abs-1807-04320,DBLP:journals/tii/LinZLPXVM18,duan2019vulsniper,zhou2019devign}.
However, these detectors still offer {\em inadequate detection capability}
and {\em inadequate locating precision}.

In order to see their {\em inadequate detection capability}, we observe that the state-of-the-art detector \cite{SySeVR}, despite its improvement upon \cite{vuldeepecker},
achieves an F1-measure of 86.0\%, a false-positive rate of 10.1\%, and a false-negative rate of 12.2\% for synthetic and academic programs, and achieves an F1-measure of 70.8\%, a false-positive rate of 18.2\%, and a false-negative rate of 32.0\% for real-world programs (see Table \ref{Table_Comparison_with_other_tools_detection} in Section \ref{subsec:Experiments_RQ3}).
This unsatisfactory performance can be attributed to their incapability in (i) capturing the relations between semantically-related statements across program {\em files} and (ii) accommodating accurate control flows and variable {\em define-use} relations. 
Note that (i) is important because programs often contain many user-defined and/or system header files (e.g., {\tt .h}) for specifying {\em types} and {\em macros}, but (i) cannot be achieved by analyzing each source code file alone because these types and macros are used in {\em program} files (e.g., {\tt .c}) while defined in {\em header} files. This mandates cross-file dependence analysis, which was not clear until now.
Note that (ii) cannot be achieved by source code analysis because it cannot accurately identify control flows and variable {\em define-use} relations \cite{SlicingLLVMBitcode}, 
which is caused by the fact that each variable is not assigned exactly once and source code-based representations have many constructs (e.g., identifiers).
As a consequence of not being able to achieve the preceding (i) and (ii), the semantics-based statements that are used for learning vulnerability detectors cannot contain enough semantic information, causing ineffective vulnerability detectors.

To see their {\em inadequate locating precision}, we observe that although they operate on program slices
(which are finer-grained than functions), a program slice can have many lines of code.
For example, according to the dataset published in \cite{SySeVR}, 78.7\% of their program slices have at least 10 lines of code and 47.8\% of them have at least 20 lines, indicating a low locating precision.
That is, coarse-grained vulnerability detection is merely a pre-step for vulnerability assessment because it cannot precisely pinpoint vulnerabilities \cite{DBLP:conf/icse/Du0LGZLJ19}.

\smallskip

\noindent{\bf Our contributions}. 
In this paper we propose a deep learning-based vulnerability detector for C programs with source code, dubbed {\em \underline{Vul}nerability \underline{Dee}p learning-based \underline{Locator}} (VulDeeLocator).
When compared with the state-of-the-art detector \cite{SySeVR}, VulDeeLocator offers, on average, (i) a 9.8\%, 7.9\%, and 8.2\% improvement in the vulnerability detection F1-measure, false-positive rate, and false-negative rate respectively, and (ii) a 4.2X improvement in the vulnerability locating precision.
When applied to 200 files randomly selected from three real-world software products (i.e., FFmpeg 2.8.2, Wireshark 2.0.5, and Libav 9.10), VulDeeLocator detects 18 confirmed vulnerabilities (i.e., true-positives). Among them, 16 vulnerabilities correspond to known vulnerabilities; the other two are not reported in the {\em National Vulnerability Database} (NVD) \cite{NVD} but have been ``silently'' patched by the vendor of Libav when releasing newer versions.
The innovations behind VulDeeLocator are in three-fold.

First, we identify one root cause of the aforementioned inadequate detection capability of existing deep learning-based vulnerability detectors, 
and address this inadequate detection capability by linking multiple files through define-use relations and leveraging intermediate code-based representations. The insight behind this approach is that intermediate code-based representations are in the {\em Static Single Assignment} (SSA) form and therefore can assure that each variable is defined-and-then-used and is assigned exactly once \cite{DBLP:books/mk/Muchnick1997}.

Second, we propose the notion of {\em granularity refinement} to locate the vulnerable lines of code.
This principle guides us to propose a specific granularity refinement method, dubbed {\em Bidirectional Recurrent Neural Network} (\underline{BRNN}) for \underline{v}ulnerability \underline{d}etection and \underline{l}ocating'' (or BRNN-vdl for short). Although this specific method is unlikely optimal, it is effective by making VulDeeLcoator output vulnerabilities about 3 lines of code; by contrast, the input program slices to VulDeeLocator are for example
32 lines of code.

Third, we prepare a vulnerability dataset in the {\em Lower Level Virtual Machine} (LLVM)
intermediate code with accompanying program source code.
This dataset is motivated by the need of evaluating the effectiveness of VulDeeLocator; it contains 157,692  vulnerability candidates in intermediate code, among which 40,450 are vulnerable and 117,242 are not vulnerable. 
It is not trivial to prepare this dataset because we need user-defined and system header files for generating intermediate code.
In order for other researchers to use the dataset, we have made the dataset and the source code used in our experiments available at \url{https://github.com/VulDeeLocator/VulDeeLocator}. 

\smallskip

\noindent{\bf Paper organization}.
Section \ref{sec:Background} discusses the basic ideas underlying VulDeeLocator.
Section \ref{sec:Design} presents an overview of VulDeeLocator.
Section \ref{sec:Design-IntermediateCode} describes how VulDeeLocator leverages intermediate code and Section \ref{sec:Design-Refinement} describes how VulDeeLocator pinpoints  vulnerabilities.
Section \ref{sec:experiments-and-results} presents our experiments and results.
Section \ref{sec:Limitations} discusses limitations of the present study.
Section \ref{sec:Related_work} reviews the related prior work.
Section \ref{sec:Conclusion} concludes the present paper.

\section{Basic Ideas}
\label{sec:Background}
\input{background}

\section{Overview of VulDeeLocator}
\label{sec:Design}
\input{design-Overview}

\section{Intermediate Code-based Vulnerability Candidate Representation}
\label{sec:Design-IntermediateCode}
\input{design-IntermediateCode}

\section{Fine-grained Vulnerability Detection}
\label{sec:Design-Refinement}
\input{design_Refinement}

\section{Experiments and Results}
\label{sec:experiments-and-results}

\input{experiments}

\section{Limitations}
\label{sec:Limitations}
\input{limitations}

\section{Related Work}
\label{sec:Related_work}
\input{related_work}

\section{Conclusion}
\label{sec:Conclusion}
We explicitly articulated two requirements for vulnerability detectors: simultaneously achieving high locating precision and high detection capability. We presented VulDeeLocator as the first deep learning-based vulnerability detector that can satisfy these two requirements when detecting vulnerabilities in C programs. We overcame two technical challenges --- capturing semantic information in programs (e.g. relations between the definitions of types and macros and their uses across files, and control flows and variable define-use relations) --- by introducing the idea of granularity refinement and leveraging the intermediate code-based representations.
As one application, VulDeeLocator detected four vulnerabilities that were {\em not} reported in the NVD.
The limitations of the present study offer interesting open problems for future research.

\section*{Acknowledgment}
We thank the reviewers for their constructive comments, which have guided us in improving the paper. 
We thank Jing Tang for collecting the real-world vulnerable programs and their patches. 
The authors from Huazhong University of Science and Technology and Hebei University were supported in part by the National Natural Science Foundation of China under Grant No. 61802106 and in part by the Natural Science Foundation of Hebei Province under Grant No. F2020201016. 
S. Xu was supported in part by ARO Grant \#W911NF-17-1-0566 as well as NSF Grants \#2122631 (previously \#1814825) and \#1736209. Any opinions, findings, conclusions or recommendations expressed in this work are those of the authors and do not reflect the views of the funding agencies in any sense.

%


\ifCLASSOPTIONcaptionsoff
  \newpage
\fi



\bibliographystyle{IEEEtran}
\bibliography{bibliography}

\appendix
\label{appendix}
We propose Algorithm \ref{alg_obtaining_ SeVC} to automatically generate iSeVCs from sSyVCs and the intermediate code. 
The algorithm has three components:
generating linked IR files (Lines \ref{algorithm:begin-step1}-\ref{algorithm:end-step1}), generating IR slices corresponding to sSyVCs (Lines \ref{algorithm:begin-step2}-\ref{algorithm:end-step2}), and generating iSeVCs (Lines \ref{algorithm:begin-step3}-\ref{algorithm:end-step3}).

\begin{algorithm}[!h]
\footnotesize
\caption{Generating iSeVCs from the intermediate code}
\label{alg_obtaining_ SeVC}
\begin{basedescript}{\desclabelstyle{\pushlabel}\desclabelwidth{6em}}
\item[Input:]
A source program $P=\{p_1, \ldots, p_n\}$, where $p_j$ ($1 \leq j \leq n$) is a source program file; a set $Y=\{y_i\}$ of sSyVCs
\item[Output:]
The set of iSeVCs $E$
\end{basedescript}
\begin{algorithmic}[1]
\STATE $E \leftarrow \emptyset$;  \COMMENT{the set of iSeVCs}
\STATE $P' \leftarrow \emptyset$; \COMMENT{the set of IR files corresponding to source program files in $P$}
\STATE $B \leftarrow \emptyset$;  \COMMENT{the set of linked IR files} for $P$

\FOR{each $p_j \in P$} \label{algorithm:begin-step1}
    \STATE Compile $p_j$ to an IR file $p_j'$;
    \STATE $P' \leftarrow P' \cup \{p_j'\}$;
\ENDFOR
\STATE Group the IR files in $P'$ by dependency relationships;
\FOR{each group $G_\mu$}
    \STATE Link the IR files in $G_\mu$ to an IR file $b_\mu'$;
    \STATE $B \leftarrow B \cup \{b_\mu'\}$;
\ENDFOR   \label{algorithm:end-step1}

\FOR{each $y_i \in Y$} \label{algorithm:begin-step2}
    \FOR{each $b_\mu' \in B$}
        \IF{the IR file corresponding to $p_j$ is linked to $b_\mu'$}
            \STATE Generate the IR slice $e_i$ corresponding to $y_i$ from $b_\mu'$;
            \STATE $E \leftarrow E \cup \{e_i\}$;
        \ENDIF
    \ENDFOR
\ENDFOR \label{algorithm:end-step2}

\FOR{each $e_i \in E$} \label{algorithm:begin-step3}
    \FOR{each function $f_\gamma$ called by function $f_\alpha$}
        \STATE The statements in $e_i$ of $f_\gamma$ are appended to the statement (in $f_\alpha$) calling function $f_\gamma$;
        \STATE Modify each numeric value of local variable
        in the appended statements to a new numeric value that has not been used in $f_\alpha$;
    \ENDFOR
\ENDFOR \label{algorithm:end-step3}

\RETURN $E$;
\end{algorithmic}
\end{algorithm}

%
%
%

%

\end{document}

%% file: background.tex
\begin{figure}[!htbp]
\centering
\includegraphics[width=.48\textwidth]{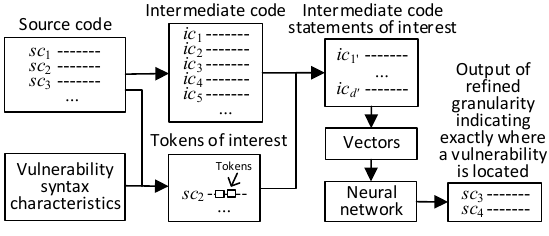}
\vspace{-0.2cm}
\caption{Illustration of VulDeeLocator, where a dashed line represents a statement containing multiple tokens.}
\label{Fig_Basic_idea}
\vspace{-0.2cm}
\end{figure}

The basic idea behind VulDeeLocator is to take advantage the best of both program analysis and deep learning techniques: generate vulnerability candidates by leveraging program analysis techniques, and use deep learning techniques to ``eliminate'' the false positives incurred by program analysis techniques. 
Specifically, VulDeeLocator extracts some {\em tokens} (e.g., identifiers, operators, constants, and keywords)
from program source code according to a given set of vulnerability syntax characteristics, and then leverage the intermediate code of the same program to accommodate the
statements in the intermediate code that are semantically related to those tokens. These statements are encoded into vectors (which are then used to train a neural network)
or are the input to the trained neural network
for vulnerability detection. The output in the testing phase is finer-grained (i.e., shorter or smaller) than the corresponding input. Fig. \ref{Fig_Basic_idea} illustrates these basic ideas, showing that in the testing phase an input of $d'$ intermediate code statements leads to a {\em refined} output of two source code statements indicating where the vulnerability is.

\begin{figure}[!htbp]
\vspace{-0.2cm}
\centering
\includegraphics[width=.48\textwidth]{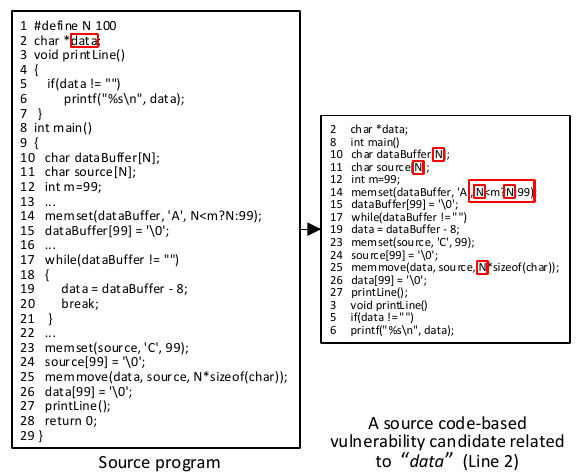}
\vspace{-0.2cm}
\caption{An example showing that any detector that cannot accommodate the semantic information related to ``{\em data}'' (Line 2 in the source program) may not detect the vulnerability.}
\label{Fig_SeVC_problems}
\vspace{-0.2cm}
\end{figure}

Fig. \ref{Fig_SeVC_problems} presents an example of buffer underflow vulnerability related to pointer ``$data$'' (Line 2).
This vulnerability may not be detected by any detector that cannot accommodate the control flow of conditional operator ``{\em N$<$m?N:99}'' in statement ``{\em  memset(dataBuffer, 'A', N$<$m?N:99);}'' (Line 14), 
because (i) the control flow is implicit in ``{\em N$<$m?N:99}'' 
rather than expressed as two branches of comparison between {\em N} and {\em m},
and (ii) the macro definition identifier {\em N} (highlighted with boxes) is not identified as 100. 
This explains why VulDeeLocator transforms ``{\em  memset(dataBuffer, 'A', N$<$m?N:99);}'' to 4  statements (Lines 6-9 in Fig. \ref{Fig_SeVC_generation}(d)) in the intermediate code-based vulnerability candidate, which 
can recognize that  {\em N} takes the value 100 and can generate two branches by comparing {\em N} and {\em m}.
Moreover, VulDeeLocator uses BRNN-vdl, which will be presented later, to pinpoint that the vulnerable line is in Line 25. 

%% file: design-Overview.tex
It is intuitive that vulnerabilities exhibit some syntax characteristics that can be leveraged to identify some pieces of code (i.e., program slices) as {\em initial candidates} for vulnerability detection \cite{vuldeepecker,SySeVR}.
Vulnerability syntax characteristics can be represented in some appropriate data structures (e.g., Abstract Syntax Trees or ASTs), 	which allow one to extract pieces of source code that match these characteristics.
These pieces of source code are the starting point for vulnerability detection,
leading to:
	
	\begin{definition}
	{\rm (\underline{s}ource code- and \underline{Sy}ntax-based \underline{V}ulnerability \underline{C}andidate or sSyVC \cite{SySeVR})}
	\label{definition:sSyVC}
	Given a source program $P$ and a set of vulnerability syntax characteristics $H=\{h_1,\ldots,h_\eta\}$, 
	an sSyVC $y_i$ is one or multiple consecutive tokens (e.g., identifiers, operators, constants, and keywords) in $P$
    that match some vulnerability syntax characteristic $h_q$ ($1 \leq q \leq \eta$).
    \end{definition}

	Given sSyVCs extracted from a source program, we propose leveraging program intermediate code to capture semantic information, leading to:
	
	\begin{definition}
		{\rm (\underline{i}ntermediate code- and \underline{Se}mantics-based \underline{V}ul\-nerability \underline{C}andidate or iSeVC)}
		\label{definition:iSeVC}
		Given a source program $P$, its intermediate code $P'$, and an sSyVC $y_i$ of $P$, 
		denote by $y_i'$ the intermediate code of sSyVC $y_i$.
		The iSeVC corresponding to sSyVC $y_i$, denoted by $e_i$, is a sequence of statements in intermediate code $P'$; these statements are data or control dependent \cite{DBLP:journals/jpl/Tip95} on $y_i'$.
		That is, the iSeVC corresponding to sSyVC $y_i$ is a program slice of $y_i'$ in the intermediate code of program $P$.

	\end{definition}

\begin{figure*}[!htbp]
\vspace{-0.2cm}
\centering
\includegraphics[width=0.85\textwidth]{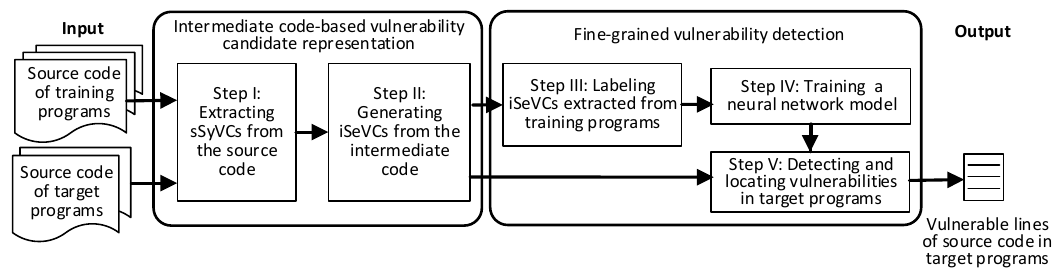}
\vspace{-0.2cm}
\caption{Overview of VulDeeLocator with two components: {\em intermediate code-based vulnerability candidate representation} and {\em fine-grained vulnerability detection}.
The {\em learning} phase consists of Steps I-IV and the {\em testing} (i.e., detection) phase consists of Steps I, II, and V.
}
\label{Fig_Overview_of_VulDeeLocator}
\vspace{-0.2cm}
\end{figure*}

Fig. \ref{Fig_Overview_of_VulDeeLocator} highlights
the structure of VulDeeLocator, which can be instantiated with specific intermediate code representations and deep learning models.
The input to VulDeeLocator is the source code of training programs for learning a neural network or target programs for vulnerability detection.
More specifically, the learning-phase input includes source code of C programs, which may or may not be vulnerable. The source code of C programs should satisfy the following: (i) they can be compiled into (platform-independent) intermediate code, such as the LLVM intermediate code \cite{DBLP:conf/cgo/LattnerA04};
and (ii) the vulnerable programs are accompanied by descriptions on the locations of their vulnerabilities, which will be leveraged to locate vulnerabilities in target programs.

At a high level, VulDeeLocator has two components. The first component leverages intermediate code representation of training programs and target programs as follows:
\begin{itemize}
\setlength{\itemsep}{0pt}
\setlength{\parsep}{0pt}
\setlength{\parskip}{2pt}
\item Step I: Extracting sSyVCs from the source code, namely pieces of code that bear some vulnerability syntax characteristic(s).
\item Step II: Generating iSeVCs from the intermediate code according to sSyVCs.
\end{itemize}

The second component uses the intermediate code-based representation to detect and locate vulnerabilities as follows:

\begin{itemize}
\setlength{\itemsep}{0pt}
\setlength{\parsep}{0pt}
\setlength{\parskip}{2pt}
\item Step III: Labeling iSeVCs extracted from training programs as vulnerable or not and vulnerability locations.
\item Step IV: Training a neural network model from the vector representations of the iSeVCs and their labels.
\item Step V:
Using the trained neural network model to detect and locate vulnerabilities in target programs. 
\end{itemize}
The {\em learning} phase corresponds to Steps I-IV and the {\em testing} (i.e., detection)
phase corresponds to Steps I, II, and V.

%% file: design-IntermediateCode.tex
\subsection{Guiding Principles for Vulnerability Candidate Representation}

It is intuitive that vulnerability detectors should accommodate program semantic information, highlighting the importance of identifying effective vulnerability candidate representations. For this purpose, we propose using the following principles to
guide the identification of effective vulnerability candidate representation. 
This paradigm of ``problem $\rightarrow$ principles (as strategies to solve the problem) $\rightarrow$ technical means (whose discovery is guided by the principles)'' is both beautiful and useful.

\begin{itemize}
\setlength{\itemsep}{0pt}
\setlength{\parsep}{0pt}
\setlength{\parskip}{3pt}
\item {\bf Principle 1: Accommodating semantically-related program statements across files}.
Some files may be dependent on others because, for example, a variable used or referred in one file may be defined in another file. Effective vulnerability candidate representations should accommodate this define-use relation.

\item {\bf Principle 2: Accommodating semantically-related program statements across functions}.
Semantically-related statements may go beyond boundaries of functions, meaning that effective vulnerability candidate representations should accommodate, and further preserve the order of, those semantically-related statements, even if they belong to different functions.
\end{itemize}

\subsection{Extracting sSyVCs}
\label{subsec:ExtractingsSyVCs}
As defined above, an sSyVC is a piece of code that is extracted from a program according to some vulnerability syntax characteristic(s).
There may be many approaches to obtaining vulnerability syntax characteristics for vulnerability detection. As a concrete example,
we leverage the syntax characteristics of known vulnerabilities and represent these characteristics via {\em Abstract Syntax Trees} (ASTs) of the program source code (more precisely, attributes of
the nodes on ASTs). This will ease the extraction of sSyVCs according to vulnerability syntax characteristics.
We define the following four kinds of vulnerability syntax characteristics, which are mentioned here because they will be referred in our examples.

\begin{itemize}
\setlength{\itemsep}{0pt}
\setlength{\parsep}{0pt}
\setlength{\parskip}{3pt}
\item {\em Library/API Function Call} (FC): This vulnerability syntax characteristic is that the type of a node on the AST is function call, the function name matches a library/API function name, and at least one argument of the function call is a variable.
\item {\em Array Definition} (AD): This vulnerability syntax characteristic is that the type of a node on the AST is variable declaration and the code corresponding  to the node contains characters `[' and `]'.
\item {\em Pointer Definition} (PD): This vulnerability syntax characteristic is that the type of a node on the AST is variable declaration and the code corresponding to the node contains character `$\ast$'.
\item {\em Arithmetic Expression} (AE): This vulnerability syntax characteristic is that the type of a node on the AST is assignment expression and the node has at least one
variable
at the right-hand side of the assignment expression.
\end{itemize}

Given the source code of a program, one can generate its AST(s), from which sSyVCs can be extracted by identifying the nodes whose type and code match some vulnerability syntax characteristics.
We reiterate that these syntax characteristics themselves are far from adequate in detecting vulnerabilities because they cannot accommodate the due semantic information that is related to vulnerabilities.

\begin{figure*}[!htbp]
	\vspace{-0.2cm}
	\centering
	\includegraphics[width=.96\textwidth]{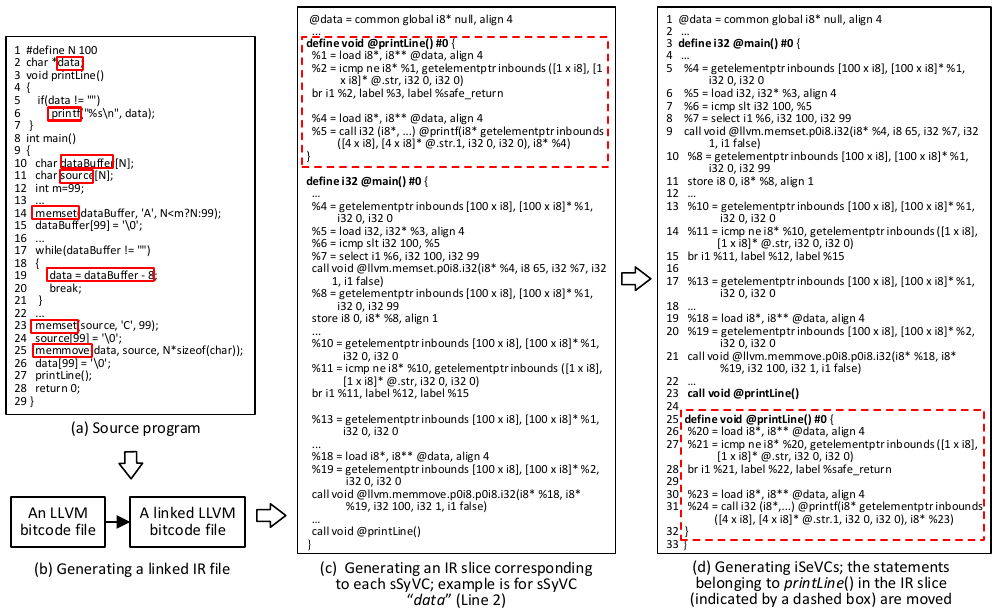}
	\vspace{-0.2cm}
	\caption{(a): An example showing the sSyVCs (highlighted by boxes) that are extracted from the program in Fig. \ref{Fig_SeVC_problems};
		(b)-(d): An example showing the generation of iSeVCs from the sSyVC ``{\em data}'' (Line 2) using the LLVM intermediate code.
	}
	\label{Fig_SeVC_generation}
	\vspace{-0.4cm}
\end{figure*}

Fig. \ref{Fig_SeVC_generation}(a) is an example showing the sSyVCs (highlighted by boxes) in a program:
sSyVCs related to the FC-kind vulnerability syntax characteristics include ``{\em printf}'' (Line 6), ``{\em memset}'' (Lines 14 and 23), and ``{\em memmove}'' (Line 25); sSyVCs related to the AD-kind vulnerability syntax characteristics include ``{\em dataBuffer}'' (Line 10) and ``{\em source}'' (Line 11); sSyVCs related to the PD-kind vulnerability syntax characteristics include ``{\em data}'' (Line 2); and sSyVCs related to the AE-kind vulnerability syntax characteristics include  ``{\em data=dataBuffer-8}'' (Line 19).

\subsection{Generating iSeVCs} 
\label{subsec:generate_IR}
Corresponding to the aforementioned principles,
there are three components for generating iSeVCs:
generating linked {\em Intermediate Representation} (IR) files; generating IR slices corresponding to sSyVCs; and generating iSeVCs. Algorithm 1 
in Appendix provides the details on generating iSeVCs.

\smallskip
\noindent{\bf Generating linked IR files (enforcing Principle 1).}
This component generates one or multiple linked IR files from source programs as follows:  
(i) use a compiler (e.g., {\em Clang}) to generate an IR file for each source file; (ii) link the IR files according to their dependence relations, leading to one or multiple linked IR files.
Fig. \ref{Fig_SeVC_generation}(b) illustrates the idea for the LLVM intermediate code using the example sSyVC ``{\em data}'', which belongs to  Line 2 of the source program described in Fig. \ref{Fig_SeVC_generation}(a). We use the LLVM intermediate code, or more specifically LLVM IR \cite{DBLP:conf/cgo/LattnerA04} as an example, because it is widely used for C programs. Specifically, we use the {\em Clang} compiler to generate LLVM bitcode files, then link them according to their dependence relations.

\smallskip
\noindent{\bf Generating IR slices corresponding to sSyVCs and generating iSeVCs (enforcing Principle 2).
}
Given a sSyVC,
we can generate a corresponding IR slice as follows: 
(i) generate a dependence graph by extracting the control and data dependencies from the linked IR file; (ii) slice the dependence graph according to each sSyVC, which can be done 
by tools such as  {\em dg} \cite{DG}. 
Fig. \ref{Fig_SeVC_generation}(c) depicts the LLVM IR slice corresponding to the aforementioned sSyVC ``{\em data}''. Each local variable is represented as a numeric value with a prefix \%; for each function in the LLVM IR slice, the numeric value is 1 for the first local variable and then increases by 1 for each subsequent local variable.

Given the IR slices, we generate iSeVCs as follows.  
For each function $f_\gamma$ that is called by function $f_\alpha$, the statements in the IR slice of function $f_\gamma$ are appended to the statement (in function $f_\alpha$) that  calls function $f_\gamma$. This is to preserve the order of statements that possibly belong to different functions but are related to each other (according to a control and/or data dependence). 
If there is a loop in the sequence of function calls (e.g., $f_\gamma$ calls $f_\alpha$ and then $f_\alpha$ calls $f_\gamma$, then $f_\gamma$ calls $f_\alpha$ and then $f_\alpha$ calls $f_\gamma$, and so on), we only consider the first loop (i.e., $f_\gamma$ calls $f_\alpha$ and then $f_\alpha$ calls $f_\gamma$ but not any further) so as to avoid an infinite loop. 
In order to avoid assigning the same numeric value to different local variables in the IR slices of different functions, each numeric value of local variable in the appended statements is modified to a new numeric value that has not been assigned.
Fig. \ref{Fig_SeVC_generation}(d) illustrates that the statements in the LLVM IR slice of function {\em printLine}, which is highlighted by the dashed box, are appended to the statement ``{\em call void @printLine()}'' in the calling function {\em main}.
The local variable ``\%1'' in the LLVM IR slice of function {\em printLine} is modified to ``\%20'' because ``19'' is the last assigned numeric value in function {\em main}, which is shown in Fig. \ref{Fig_SeVC_generation}(c).

%% file: design_Refinement.tex
\subsection{Requirements for Fine-grained Vulnerability Detectors}
We propose the following three requirements for neural network models that aim to detect and locate vulnerabilities. 

\begin{itemize}
\item {\bf Requirement 1:  Granularity refinement}.
The granularity of code determines the unit of source code for analysis, which can range from the coarsest granularity in component, to file, to function, to code fragment, to statement, and to the finest granularity in token. Granularity refinement is critical to pin down vulnerabilities or precisely identify the vulnerable lines of code, which corresponds to the granularity in {\em statement}. This is so because the input to the vulnerability detector is an iSeVC, which corresponds to the coarse granularity in {\em code fragment}, meaning that the granularity of the output of the vulnerability detector is finer than that of the input.

\item {\bf Requirement 2: Easy mapping}.
It should be easy to map the output of a neural network (at a refined granularity) back to the iSeVCs to pinpoint vulnerabilities.
The output should be a sequence of tokens, where one or multiple consecutive tokens correspond to a same line of code in the intermediate code.
These lines of intermediate code can be easily mapped back to iSeVCs, and therefore the
vulnerable lines of code in source programs.

\item {\bf Requirement 3: Attention taking}.
The notion of attention is borrowed from deep learning and corresponds to the important parts of the input a learner should focus on; technically, attention is achieved by properly assigning weights in a neural network (i.e., more attention means a higher weight). For a vulnerable iSeVC, it is likely that only one or few statements are vulnerable while the others are not, which means that the vulnerable statements are more ``important'' than the non-vulnerable ones and therefore should be given a higher weight (i.e., more attention) by a neural network. 
\end{itemize}

\subsection{Labeling iSeVCs}
We label each iSeVC from training programs as follows: If an iSeVC contains a known vulnerability, the iSeVC is labeled with the line number(s) of the vulnerability in the iSeVC (i.e. location of the vulnerability), denoted by $x_1, \ldots, x_\zeta$ where $x_\epsilon$ ($1 \leq \epsilon \leq \zeta$) is a line number corresponding to the vulnerability; otherwise, the iSeVC is labeled as ``0'' (i.e., containing no vulnerability).
Since a vulnerability dataset should provide the locations (e.g., line numbers) of vulnerabilities in the intermediate code of a source program, these line numbers of vulnerabilities in the source program need to be
mapped to the line numbers in the intermediate code, which can be done simply by leveraging a textual LLVM file that comes with debugging information.

\subsection{Training a Neural Network Model}
Each iSeVC needs to be encoded into a vector, which is used as an input to a neural network.
In order to make iSeVCs independent of user-defined function names while capturing program semantic information, this step maps user-defined function names to symbolic names (e.g., ``FUN1'', ``FUN2'') in a one-to-one fashion.
It is worth mentioning that iSeVCs are already independent of local variable names because the latter are replaced with symbolic names in the intermediate code.
Then, a {\em word embedding} method can be used to encode iSeVCs into vectors.
Since the lengths of the resulting vectors 
can be different and a neural network takes input vectors of a fixed-length $\theta$, these vectors may need to be adjusted as follows:
If a vector is shorter than $\theta$, zeroes are padded to the end of the vector; if a vector is longer than $\theta$, the vector is truncated to length $\theta$ to make the sSyVC appear in the middle of the resulting vector \cite{SySeVR}.
Finally, the vectors are used as input to a neural network that satisfies the aforementioned Requirements 1-3, as shown in Sections \ref{subsec:Experiments_RQ1}-\ref{subsec:Experiments_RQ4}. 
In what follows, we elaborate the neural network BRNN-vdl we propose, which satisfies the aforementioned Requirements 1-3.

\subsubsection{BRNNs achieve easy mapping}
One may suggest to use {\em Recurrent Neural Networks} (RNNs), such as {\em Long Short-Term Memory} (LSTM) and {\em Gated Recurrent Unit} (GRU), 
to achieve {\em easy mapping}. This seems reasonable because RNNs are effective in coping
with sequential data and the output at each time step corresponds to a token in an iSeVC, which makes it easy to map the output back to iSeVCs. 
As shown in Fig. \ref{Fig_DL_model_training}, each output of the activation layer corresponds to a token of an input iSeVC at each time step. The outputs of standard BRNN can be mapped back to the vulnerable tokens, and therefore the vulnerable lines of intermediate code (i.e., multiple continuous vulnerable tokens).  
However, unidirectional RNNs are not sufficient because a statement in a program may be affected by some {\em preceding} and/or {\em subsequent} statements in the program.
Nevertheless, BRNNs, such as {\em Bidirectional LSTM} (BLSTM) and {\em Bidirectional GRU} (BGRU) can indeed achieve {\em easy mapping}, while accommodating {\em preceding} and {\em subsequent} statements.

BRNNs cannot achieve the other two properties because their output granularity is the same as, rather than refining, the input granularity
because they treat every part of an input equally. For vulnerability detection, some parts (i.e., vulnerable lines of code) of an iSeVC may be more important than the other parts of the iSeVC and should be paid more {\em attention} by neural networks.

\subsubsection{BRNN-vdl: A novel variant of BRNN further achieving attention taking and granularity refinement}
\begin{figure}[!htbp]
\centering
\includegraphics[width=.4\textwidth]{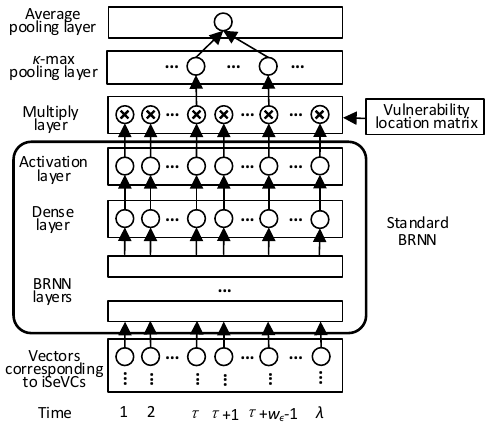}
\vspace{-0.2cm}
\caption{BRNN-vdl extends BRNN with three extra layers (i.e., the {\em multiply}, {\em $\kappa$-max pooling}, and {\em average pooling} layers) that formulate the ``vdl'' part to achieve three desired properties.
}
\label{Fig_DL_model_training}
\vspace{-0.2cm}
\end{figure}

Fig. \ref{Fig_DL_model_training} highlights the structure of BRNN-vdl, which extends the standard BRNN with three extra layers that formulate the ``vdl'' part to achieve granularity refinement and attention taking.
The input to BRNN-vdl includes (i) the vectors that represent the iSeVCs, and (ii) a vulnerability location matrix that represents the locations of vulnerabilities in each vector.
The learning phase outputs a BRNN-vdl with fine-tuned parameters.
In what follows, we briefly review BRNN and then describe the three extra layers in BRNN-vdl we introduce.

\smallskip

\noindent{\bf Overview of the BRNN component in BRNN-vdl.}
As shown in Fig. \ref{Fig_DL_model_training}, the {\em standard BRNN} has (i) a number of BRNN layers, which connect the RNN cells (e.g., LSTM and GRU) in both forward and backward directions, (ii) a dense layer, which reduces the number of dimensions of the vectors received from the BRNN layers, and (iii) an activation layer, which uses an activation function to generate the output at a time step. In the context of the present paper, the input is the vectors representing the labelled iSeVCs.
Each time step corresponds to a token in an iSeVC.
At time step $\tau$, where $1 \leq \tau \leq \lambda$ and $\lambda$ is the number of tokens in each iSeVC,
the output of the
BRNN layers for iSeVC $e_i$, denoted by $g_\tau(e_i)$, is
\begin{equation}
  g_\tau(e_i) = \phi(g_{\tau-1}(e_i),g_{\tau+1}(e_i),e_i,\bm{\omega},\bm{\beta}),
\end{equation}
where $\bm{\omega}$ is a weight vector, $\bm{\beta}$ is a bias vector, $g_{\tau-1}(e_i)$ and $g_{\tau+1}(e_i)$ are respectively the output of the
BRNN layers at time steps $\tau-1$ and $\tau+1$, and function $\phi$ indicates that the output of BRNN layers is represented by its parameters that include $g_{\tau-1}(e_i)$, $g_{\tau+1}(e_i)$, $e_i$, $\bm{\omega}$, and $\bm{\beta}$. How these parameters exactly interact with each other depends on the RNN cells, such as LSTM and GRU. 
For iSeVC $e_i$, the output vector of the standard BRNN $\bm{A_i}$ (i.e., the output vector of the activation layer) is denoted by
\begin{equation}
  \bm{A_i} = ( g_1(e_i), \ldots, g_\lambda(e_i)).
\end{equation}

\smallskip

\noindent{\bf The multiply layer achieves {\em attention taking}.}
In order to enable the neural network to predict vulnerability locations, the {\em multiply} layer needs to treat different iSeVCs differently according to the locations of the tokens corresponding to the vulnerable lines of code.
(i) For the iSeVCs that are vulnerable, the multiply layer is meant to select the outputs of the tokens that correspond to the vulnerable lines of code.
These selected outputs will be used in the subsequent layers and the back propagation process of BRNN-vdl because they would help locate vulnerabilities with a higher precision (than not using this multiply layer).
(ii) For the iSeVCs that are not vulnerable, the multiply layer is meant to select {\em all} outputs of the tokens and use them in the subsequent layers and the back propagation process of BRNN-vdl, because these tokens are equally important as far as the learning phase is concerned.

This design choice can be justified as follows: For each iSeVC that is not vulnerable, all tokens are treated as equal because no line of code is vulnerable.
However, a vulnerable iSeVC contains (i) one or multiple vulnerable lines of code, which should be highlighted for vulnerability locating purposes, and (ii) possibly a large number of lines of code that are not vulnerable, which only provide the context for vulnerability detection. If all tokens in a vulnerable iSeVC are treated equal, a false-negative can occur because most lines of code are not vulnerable.

Formally, for iSeVC $e_i$, the multiply layer multiplies the output vector of the activation layer $\bm{A_i}$ with the vulnerability location matrix $\bm{L_i}$.
The output vector of the multiply layer $\bm{M_i}$ is denoted by
\begin{equation}
   \bm{M_i}=\bm{A_{i}L_{i}},
\end{equation}
where $\bm{L_i}$ is a diagonal matrix with $\bm{L_i}=diag(\alpha_1, \alpha_2, \ldots,$ $\alpha_\lambda)$.
For a vulnerable iSeVC, let us denote by $x_\epsilon'$ the location of the first token in the vulnerable line $x_\epsilon$ for some $1 \leq \epsilon \leq \zeta$,
and by $w_\epsilon$ the number of tokens in $x_\epsilon$.
The value of $\alpha_\varphi$ ($1 \leq \varphi \leq \lambda$) is determined as follows:
For the iSeVCs that are vulnerable, if $\varphi \in \{x_\epsilon', \ldots, x_\epsilon'+w_\epsilon-1\}$, then we set $\alpha_\varphi=1$; otherwise, we set $\alpha_\varphi=0$.
For the iSeVCs that are not vulnerable,
we set $\alpha_\varphi=1$ for $1 \leq \varphi \leq \lambda$.

\smallskip

\noindent{\bf The $\kappa$-max pooling layer and the average pooling layer together achieve {\em granularity refinement}.}
In order to use back propagation to train the neural network, the $\kappa$-max pooling layer and the average pooling layer need to select and process the outputs of the multiply layer.
The {\em $\kappa$-max pooling} layer is meant to select the $\kappa$ largest values among the elements in the output vector of the multiply layer $\bm{M_i}$.
The {\em average pooling} layer is meant to compute the average of the outputs of the $\kappa$-max pooling layer.
Intuitively, these two layers together achieve granularity refinement
because (i) they further select the outputs of the multiply layer to obtain the output corresponding to each iSeVC, which is used for back propagation, and (ii) they take into account both the maximum and the average.

Formally, for an iSeVC $e_i$, the output of average pooling layer $o_i$ is defined as
\begin{equation}
   o_i=ave(max_{\kappa}(\bm{M_i})),
\end{equation}
where function $max_\kappa$ returns the $\kappa$ largest elements in the vector, and function $ave$ returns the average of the $\kappa$ largest elements.
After conducting iterative forward and backward propagations, the training process converges to a BRNN-vdl with fine-tuned parameters, which encodes vulnerability patterns in the training data.

\subsection{Detecting and Locating Vulnerabilities}
\begin{figure}[!htb]
	\vspace{-0.2cm}
	\centering
	\includegraphics[width=.32\textwidth]{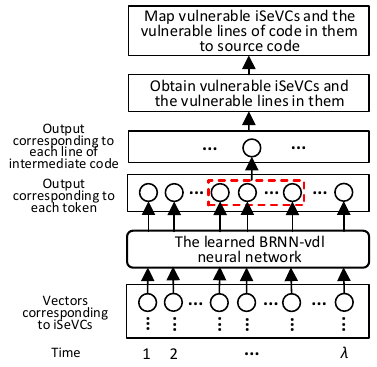}
	\vspace{-0.2cm}
	\caption{Using the learned BRNN-vdl
		to detect vulnerabilities in target programs, where the dashed box highlights the tokens extracted from a line of code.
	}
	\label{Fig_DL_model_detection}
	\vspace{-0.2cm}
\end{figure}

Fig. \ref{Fig_DL_model_detection} highlights using the learned BRNN-vdl to detect and locate vulnerabilities in target programs. The input is the vectors representing the iSeVCs extracted from the target programs.
We first obtain the outputs of the activation layer corresponding to the tokens in iSeVCs, and	compute the average of the $\kappa$ largest output values for the tokens in each line.
Then, we extract the lines of code whose output is greater than the threshold $\vartheta$, leading to vulnerable iSeVCs and vulnerable lines of code.
Finally, we map the vulnerable iSeVCs and the vulnerable lines of code in them to source code as the output of the detection phase.

Take the iSeVC in Fig. \ref{Fig_SeVC_generation}(d) as an example. The vector corresponding to iSeVC is the input to the learned BRNN-vdl neural network. After computing the average of the $\kappa$ largest output values for the tokens in each line, the iSeVC is determined as vulnerable and the vulnerable lines of intermediate code is in Lines 19-21 in Fig. \ref{Fig_SeVC_generation}(d). Finally, the vulnerable lines in iSeVC are mapped to the vulnerable line in the source code (i.e., Line 25 in Fig. \ref{Fig_SeVC_generation}(a)).

%% file: experiments.tex
We gear our experiments towards answering the following four {\em Research Questions} (RQs):
\begin{itemize}
\item RQ1: Can intermediate code-based vulnerability candidate representation be leveraged to achieve a substantially higher vulnerability detection capability?
\item RQ2: Can BRNN-vdl achieve a substantially higher vulnerability locating precision than BRNN?
\item RQ3: How effective and precise is VulDeeLocator in detecting and locating vulnerabilities of target programs with known ground truth?
\item RQ4: How effective and precise is VulDeeLocator when applied to detect and locate vulnerabilities in real-world software products? 
\end{itemize}
Our experiments use a machine with a NVIDIA GeForce GTX 1080 GPU and an Intel Xeon E5-1620 CPU operating at 3.50GHz.

\subsection{Evaluation Metrics}
\label{subsec:Experiments_metrics}

We propose using five standard metrics \cite{DBLP:journals/csur/PendletonGCX17} to measure the detection capability of a vulnerability detector. 
Let {\sf TP} denote the number of vulnerable samples that are detected as vulnerable (i.e., true-positives),
{\sf FP} denote the number of samples that are not vulnerable but are detected as vulnerable (i.e., false-positives),
{\sf TN} denote the number of samples that are not vulnerable and are not detected as vulnerable (i.e., true-negatives),
{\sf FN} denote the number of vulnerable samples that are not detected as vulnerable (i.e., false-negatives).
The five metrics are:
(i) false-positive rate $FPR=\frac{{\sf FP}}{{\sf FP}+{\sf TN}}$;
(ii) false-negative rate $FNR=\frac{{\sf FN}}{{\sf TP}+{\sf FN}}$;
(iii) accuracy $A=\frac{{\sf TP}+{\sf TN}}{{\sf TP}+{\sf FP}+{\sf TN}+{\sf FN}}$;
(iv) precision $P=\frac{{\sf TP}}{{\sf TP}+{\sf FP}}$;
(v) F1-measure $F1=\frac{2 \cdot P \cdot (1-FNR)}{P+(1-FNR)}$, or the overall effectiveness.

\begin{figure}[!htbp]
\centering
\includegraphics[width=.35\textwidth]{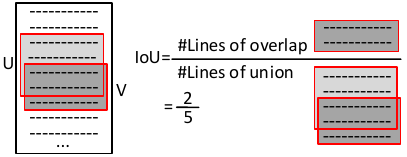}
\vspace{-0.2cm}
\caption{An example illustrating the meaning of IoU, where a dashed line represents
a program statement.
}
\label{Fig_IoU_metric}
\vspace{-0.2cm}
\end{figure}

In order to evaluate the locating precision of a vulnerability detector, we propose using the standard {\em Intersection over Union} (IoU) metric with $\text{IoU}=\frac{|{\sf U} \cap {\sf V}|}{|{\sf U} \cup {\sf V}|}$,
where {\sf U} is the set of truly vulnerable lines of code and {\sf V} is the set of detected vulnerable lines of code  \cite{IoU}. Fig. \ref{Fig_IoU_metric} illustrates the meaning of IoU with respect to one iSeVC; as highlighted by boxes, $U$ contains 4 statements,
$V$ contains 3 statements, $U\cap V$ contains 2 statements (i.e., $|U\cap V|=2$), and $U\cup V$ contains 5 statements (i.e., $|U\cup V|=5$), leading to $\text{IoU}=2/5$.
Intuitively, IoU reflects the degree at which the detected vulnerable statements overlap with the truly vulnerable statements. The closer the IoU is to 1, the higher the locating precision.

\subsection{Preparing the Input to VulDeeLocator}
We collect the source code of C programs from two vulnerability sources: the NVD \cite{NVD} and the {\em Software Assurance Reference Dataset} (SARD) \cite{SARD}.
The programs collected from the NVD are accompanied by their {\em diff} files, which describe the difference between the programs before and after patching the vulnerabilities in question.
The programs collected from the SARD are accompanied by labels, which indicate whether they are vulnerable
or not.
Note that SARD contains production, synthetic, and academic programs (i.e., test cases). We filter out the programs that cannot be compiled into the LLVM intermediate code. 

We collect 14,511 programs, including 2,182 real-world programs and 12,329 synthetic and academic programs. The real-world programs are from multiple versions of open source software written in the C language (e.g., Linux kernel, OpenSSL, FFmpeg, Wireshark, and Libtiff). We collect (i) the vulnerable programs which are reported prior to 2017 and (ii) their corresponding patched programs. The reason for collecting (i) is that we conduct experiments on real-world software products to detect vulnerabilities that are reported between 2017 and 2019, which are unknown vulnerabilities with respect to the training data.
The synthetic and academic programs are from the test cases in the SARD, where each program is labeled as good (not vulnerable), bad (vulnerable), or mixed (vulnerable functions and their patched versions provided).
We randomly choose 80\% of the synthetic or academic programs and 80\% of the real-world programs as training programs, and the remaining 20\% of the synthetic and academic programs (dubbed ``Test-set-1'') and 20\% of the real-world programs (dubbed ``Test-set-2'') as target programs. We consider these two sets of target programs because we want to see the impact of data sources.

\subsection{Intermediate Code-based Vulnerability Candidate}
\noindent{\bf Extracting sSyVCs.}
In order to extract sSyVCs from the source code, we use {\em Clang} 
to generate ASTs from a source program.
Then, we traverse the ASTs to generate sSyVCs.
For obtaining vulnerability syntax characteristics, we leverage the C vulnerability rules of the commercial
tool Checkmarx \cite{Checkmarx} because we found that the syntax characteristics from these rules have a good coverage over known vulnerabilities. This leads to the four kinds of vulnerability syntax characteristics mentioned above, namely
{\em Library/API Function Call} (FC), {\em Array Definition} (AD), {\em Pointer Definition} (PD), and {\em Arithmetic Expression} (AE). These characteristics cover 98.3\% of the vulnerable programs collected from the NVD and the SARD.

Note that we do {\em not} use the vulnerability detection results of vulnerability detector Checkmarx (i.e., output of Checkmarx) as the basis of VulDeeLocator. Instead, we only leverage the vulnerability rules of Checkmarx to extract 4 kinds of vulnerability syntax characteristics (i.e., sSyVCs). It is in this sense we generate vulnerability candidates by leveraging program analysis techniques. 
We stress that the number of vulnerability candidates that are used as input to VulDeeLocator (including both vulnerable samples and non-vulnerable samples) is much larger than the number of vulnerabilities detected by Checkmarx (i.e., vulnerable samples output by Checkmarx).

Take vulnerabilities of the FC-kind (i.e., Library/API function call) as an example, the syntax characteristic is that the type of a node on the AST in question is function call, the function name matches a library/API function name, and at least one argument of the function call is a variable.
Given the ASTs of a source program, the matching algorithm proceeds as follows:
(i) we first traverse the ASTs to identify the nodes whose type is ``CxCursor\_CallExpr'' (meaning a function call); (ii) identify the nodes whose token matches a library/API function name (e.g., memset);  (iii) traverse the children of the node corresponding to a library/API function call to identify the nodes
whose type is ``CxCursor\_DeclRefExpr'' (meaning a variable argument); and (iv) a library/API function call together with its variable arguments is extracted as an sSyVC.
In total, we extract 157,692 sSyVCs, including 40,430 sSyVCs of the FC-kind, 37,692 sSyVCs of the AD-kind, 50,266 sSyVCs of the PD-kind, and 29,304 sSyVCs of the AE-kind.

\noindent{\bf Generating iSeVCs.}
We use tool {\em dg} \cite{DG} to generate LLVM-based intermediate code slices corresponding to given source code sSyVCs as follows.
For each given source code sSyVC, the corresponding iSeVC is a slice of intermediate code statements consisting of: (i) the intermediate code statements corresponding to the given source code sSyVC; and (ii) any intermediate code statement that has a data-dependence or control-dependence with any of the variables that are used or defined in the given sSyVC. Note that the statements in (ii)  may belong to different functions or files than the one to which the given sSyVC belongs.
Take the sSyVC ``{\em data}'' (Line 2) in Fig. \ref{Fig_SeVC_generation}(a) as an example, the LLVM-based intermediate code slices (i.e., LLVM slices) corresponding to ``{\em data}'' generated by {\em dg} is described in Fig. \ref{Fig_SeVC_generation}(c).
In total, we extract 157,692 iSeVCs, including 40,450 vulnerable iSeVCs and 117,242 non-vulnerable iSeVCs.
The ratio of vulnerable to non-vulnerable iSeVCs is about 1:3.

\subsection{Fine-grained Vulnerability Detection}
\noindent{\bf Labeling iSeVCs.}
For the iSeVCs extracted from programs collected from the NVD, we focus on the vulnerabilities that are accompanied by {\em diff} files that involve {\em line deletion} or {\em line movement} because they allow us to pin down the locations of these vulnerabilities (i.e., statements prefixed by ``-'' in a {\em diff} file).
These iSeVCs are automatically labeled as follows: 
if (i) an iSeVC contains some intermediate code that corresponds to one or multiple statements prefixed by ``-'' in the {\em diff} file and (ii) the program in question contains a vulnerability, then the iSeVC is automatically labeled as the line number(s) of the vulnerability in the intermediate code;
otherwise, the iSeVC is automatically labeled as ``0'' (i.e., not vulnerable).

For iSeVCs extracted from programs collected from SARD,
if an iSeVC contains some intermediate code that corresponds to one or multiple vulnerable statements in the source program from SARD, then the iSeVC is automatically labeled as the line number(s) of the vulnerability in the intermediate code; otherwise, the iSeVC is automatically labeled as ``0'' (i.e., not vulnerable).

\noindent{\bf Training, detecting and locating.}
In order to use neural networks, we need to encode iSeVCs into vectors.
For this purpose, we first divide each iSeVC into a sequence of tokens via lexical analysis (e.g., ``call'', ``void'', ``@'', ``FUN1'', ``('', and ``)''), and then transform each token to a fixed-length vector via {\em word2vec} tool \cite{word2vec}.
Finally, a token-level vector for each iSeVC is obtained by concatenating the token-level vectors in sequence.
Each token is encoded into a vector of length 30, and each iSeVC is
represented by a vector of length $\theta$=27,000, which means that the first 900 tokens of an iSeVC are considered.

We implement the BRNN-vdl in Python using TensorFlow together with Keras. 
In order to prevent overfitting, we use dropout to ignore some units in the neural network which are chosen at random and a stratified 10-fold cross-validation to train the BRNN-vdl, while considering the trade-off between model training time and model generalization ability. We choose the hyper-parameter values that lead to the highest F1-measure. When we adjust a hyper-parameter, we set the other hyper-parameters to their default values when such default values are available, and to the values that are widely used by the deep learning community otherwise.
We implement two instances of BRNN: one is BLSTM, which leads to ``VulDeeLocator-BLSTM''; the other is BGRU, which leads to ``VulDeeLocator-BGRU''.
Take VulDeeLocator-BGRU as an example, the trained hyper-parameters are:
output dimension is 512; the number of hidden layers is 2; the number of hidden nodes at each layer is 900; batch size is 16; minibatch stochastic gradient descent together with ADAMAX 
is used; learning rate is 0.002; dropout is 0.4; the number of epochs is 10;
and $\kappa=1$.

For detecting vulnerabilities in target programs, we first compute the average of the $\kappa$ largest values among the tokens in each line of intermediate code. Then, we extract the lines whose output is larger than threshold $\vartheta$ (e.g., 0.5).
These lines of code are the vulnerable ones,
and are mapped back to the vulnerable lines of source code as the output of the test (i.e., detection) phase.

\subsection{Experiments for Answering RQ1}
\label{subsec:Experiments_RQ1}
In order to evaluate the advantages of intermediate code-based vulnerability candidate representation over source code-based one, we conduct experiments with the following two vulnerability candidate representations:

\begin{itemize}
\item {\em \underline{s}ource code- and \underline{Se}mantics-based \underline{V}ulnerability \underline{C}andidate} (sSeVC): A sSeVC is a sequence of source code statements that have some data-dependence or control-dependence with an sSyVC (i.e., source code- and Syntax-based Vulnerability Candidate, as defined in Section \ref{sec:Design}) and can be obtained by using a source code static analysis tool (e.g., {\em Joern} \cite{yamaguchi2014modeling}).
\item iSeVC: An iSeVC is a sequence of intermediate code statements that have some data-dependence or control-dependence with an sSyVC. Compared with source code-based representation, iSeVC is in the static single assignment (i.e., SSA) form, which assures that each variable is defined-and-then-used and is assigned exactly once.
\end{itemize}

We use the target programs in Test-set-1 to test 
VulDeeLocator-BGRU, while noting that experimental results with Vul\-DeeLo\-cator-BLSTM are similar. Table \ref{Table_Compilation_RQ1} summarizes the comparison. We observe that iSeVCs lead to better results than sSeVCs,
including a 4.6\% improvement in false-positive rate, a 7.4\% improvement in false-negative rate, a 6.5\% improvement in accuracy, a 5.9\% improvement in precision, and a 6.7\% improvement in F1-measure. This can be attributed to the two advantages of intermediate code-based representation:
(i) intermediate code is in the SSA form, which can expose more information about control-flows and the define-use relations between variables; (ii) intermediate code-based vulnerability candidates can capture more semantic information (e.g. the relations between the definitions of types or macros and their uses), which however may not be identified by sSeVCs. This is justified by the following two examples.

\begin{table}[!htbp]
\vspace{-0.2cm}
\caption{Vulnerability detection capability of VulDeeLocator-BGRU using two different kinds of vulnerability candidate representations
 }
\vspace{-0.2cm}
\label{Table_Compilation_RQ1}
\footnotesize
\centering
\begin{tabular}{|c|c|c|c|c|c|c|}
\hline
{\tabincell{c}{Vulnerability \\candidate}} & Representation  & \tabincell{c}{FPR \\(\%)} & \tabincell{c}{FNR \\(\%)} & \tabincell{c}{A \\(\%)} & \tabincell{c}{P \\(\%)} & \tabincell{c}{F1 \\(\%)} \\
\hline
{\tabincell{c}{sSeVC}} & \tabincell{c}{Source \\code-based}  & 5.1 & 11.7 & 92.2 & 92.3 & 90.2\\
\hline
{\tabincell{c}{iSeVC}} & \tabincell{c}{Intermediate \\code-based}  & 0.5 & 4.3 & 98.7 & 98.2 & 96.9 \\
\hline
\end{tabular}
\vspace{-0.2cm}
\end{table}

\begin{figure}[!htb]
\centering
\includegraphics[width=.48\textwidth]{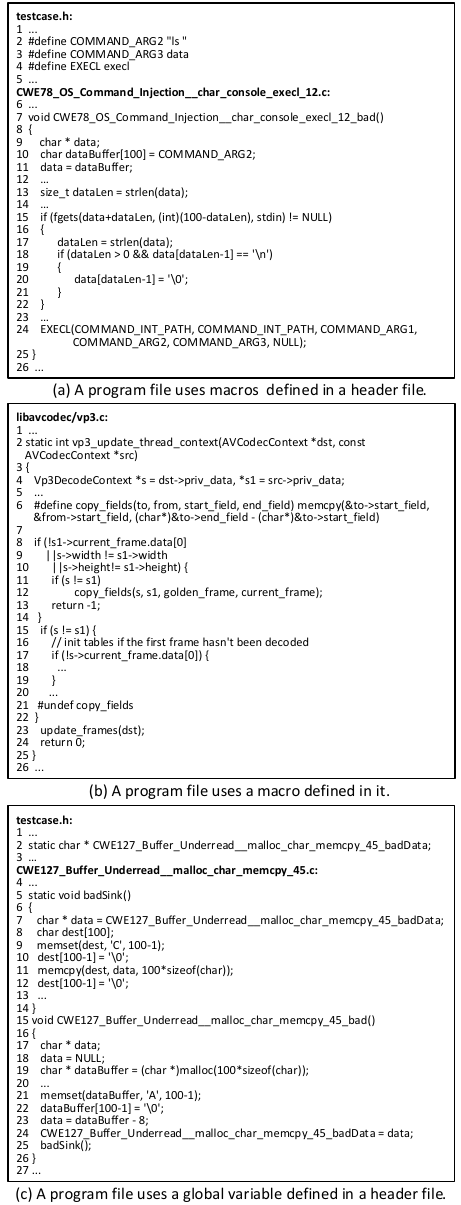}
\vspace{-0.2cm}
\caption{Three vulnerabilities that are missed by VulDeeLocator-BGRU trained from sSeVCs.}
\label{Fig_SARD_test_cases}
\vspace{-0.2cm}
\end{figure}

Fig. \ref{Fig_SARD_test_cases}(a) describes an example  macro, which contains a command injection vulnerability because the input is received from the console and is used without validation (vulnerable Line 24).
Consider the sSyVC ``$data$'' in Line 9. A source code parser (e.g., {\em Joern} \cite{yamaguchi2014modeling}) is not able to identify this macro because it is used in a program file (i.e., {\tt .c}) but  defined in a header file, while recalling that program files and header files are linked after complication.  
That is, ``COMMAND\_ARG3'' in Line 24 cannot be identified as ``$data$''. As a consequence, the corresponding sSeVC fails to identify the vulnerable statement in Line 24. This explains the false-negative.
On the other hand, the iSeVCs can identify the ``COMMAND\_ARG3'' in Line 24 as ``$data$'' after compilation.
This explains why the resulting model can detect the vulnerability.

Fig. \ref{Fig_SARD_test_cases}(b) presents another example of macro, which corresponds to vulnerability CVE-2011-3934 in FFmpeg 0.9.4.
This vulnerability is related to {\em double free} because the statement in Line 12 can cause a double release of variable ``$s \rightarrow current\_frame$''.
Consider the sSyVC ``$s$'' in Line 4. 
Although there are no relations between the semantically-related statements across two files, a source code parser (e.g., {\em Joern} \cite{yamaguchi2014modeling}) is not able to identify the complicated macro  ``$copy\_fields$($s$, $s1$, $golden\_frame$, $current\_frame)$'' in Line 12  as ``$memcpy$($\&s$ $\to$ $golden\_frame$, $\&s1$ $\to$ $golden\_frame$, $(char*)\&s$ $\to$ $current\_frame$ - $(char*)\&s$ $\to$ $golden\_frame)$,'' causing a detector not able to identify the vulnerable statement in Line 12. 
On the other hand, the iSeVC can convert the statement in Line 12 to ``$memcpy$($\&s \to golden\_frame$, $\&s1 \to golden\_frame$, $(char\ast)\&s \to current\_frame$ - $(char\ast)\&s \to golden\_frame)$;''.
This explains why a detector leveraging intermediate code can detect this vulnerability.

Fig. \ref{Fig_SARD_test_cases}(c) shows an example using a global variable,
which contains a buffer under-read vulnerability because the copy from a memory location may be located before the source buffer (vulnerable Line 11).
Consider sSyVC ``$data$'' in Line 7. 
A source code parser (e.g. {\em Joern} \cite{yamaguchi2014modeling}) is not able to identify the relationship between the definition of a global variable in a header file and its usages in different 
program files (e.g., {\tt .c}),
because program files and header files are usually linked after compilation. 
As a consequence, the sSeVC corresponding to sSyVC ``$data$'' in Line 7 does not contain any statement that is semantically related to sSyVC ``$data$'' via global variable {\em CWE127\_Buffer\_Underread\_\_malloc\_char \_memcpy\_45\_badData} in function {\em CWE127\_Buffer\_Un\-derread\_\_malloc\_char\_memcpy\_45\_bad}. 
The root cause of the vulnerability is that the data pointer points to a memory address that is different from the allocated memory buffer (Line 23), which is defined in function {\em CWE127\_Buffer\_Underread\_\_malloc\_char\_memcpy\_45\_bad}. This explains why the vulnerability is missed.
However, the model learned from iSeVCs can identify and accommodate these statements because they are semantically related to the global variable.

\smallskip
\noindent{\bf Impact of imbalanced data processing.}
Because of the 1:3 imbalance of vulnerable to non-vulnerable iSeVCs, we use the under-sampling method NearMiss-2 \cite{mani2003knn} and the over-sampling method SMOTE \cite{chawla2002smote} to evaluate the impact of imbalanced data processing. 
For this purpose,
we need to extend the five metrics defined in Section \ref{subsec:Experiments_metrics} for vulnerable class to non-vulnerable class, denoted by FPR', FNR', A', P', and F1', respectively.

\begin{table}[!htbp]
\vspace{-0.2cm}
\caption{The effectiveness of VulDeeLocator-BGRU with and without imbalanced data processing for vulnerable and non-vulnerable classes
 }
\vspace{-0.2cm}
\label{Table_imbalance}
\scriptsize
\centering
\begin{tabular}{|c|c|c|c|c|}
\hline
 & \tabincell{c}{Imbalanced data \\processing method} & None & NearMiss-2 & SMOTE \\
\hline
{\multirow{5}{*}{\tabincell{c}{Vulnerable\\ class}}} & FPR (\%) & 0.5 & 30.3 & 0.5\\
\cline{2-5}
& FNR (\%) & 4.3 & 15.7 & 3.8\\
\cline{2-5}
& A (\%) & 98.7 & 72.9 & 98.8\\
\cline{2-5}
& P (\%) & 98.2 & 43.6 & 98.2\\
\cline{2-5}
& F1 (\%) & 96.9 & 57.5 & 97.2\\
\hline
{\multirow{5}{*}{\tabincell{c}{Non-vulnerable\\ class}}} & FPR' (\%) & 4.3 & 15.7 & 3.8 \\
\cline{2-5}
& FNR' (\%) & 0.5 & 30.3 & 0.5 \\
\cline{2-5}
& A' (\%) & 98.7 & 72.9 & 98.8\\
\cline{2-5}
& P' (\%) & 98.8 & 94.1 & 98.9\\
\cline{2-5}
& F1' (\%) & 99.2 & 80.1 & 99.2\\
\hline
\end{tabular}
\vspace{-0.2cm}
\end{table}

Table \ref{Table_imbalance} shows the effectiveness of VulDeeLocator-BGRU (using iSeVC) with and without imbalanced data processing for vulnerable and non-vulnerable classes. For the vulnerable class, we observe that under-sampling incurs (for example) a 39.4\% lower F1-measure when compared with no imbalanced data processing (``None'').
This means that under-sampling non-vulnerable iSeVCs substantially decreases the detector's effectiveness because of 
the reduction in data.
Over-sampling the vulnerable class incurs (for example) a 0.3\% higher F1-measure when compared with no imbalanced data processing, at the price of a higher training time (83,216.2 vs. 51,602.8  seconds).
For the non-vulnerable class, we observe a similar phenomenon. 
The non-vulnerable class does not achieve a significantly higher F1-measure than the vulnerable class in the absence of imbalanced data processing, meaning that the model is not obviously biased towards the non-vulnerable class despite that it has more iSeVCs.
In summary, the imbalanced data processing techniques we experimented with are not significant in this specific setting, perhaps because the ratio 1:3 is not severe enough.

\smallskip
\noindent{\bf Impact of selecting training set.}
In order to determine whether different selections of training set would impact the evaluation result, we randomly select 80\% of the synthetic and academic programs and 80\% of the real-world programs as training programs for 5 times, and perform a stratified 10-fold cross-validation to train VulDeeLocator-BGRU (using iSeVC). In each 10-fold cross-validation, we obtain 10 values of FPR, FNR, accuracy, precision, and F1 corresponding to the 10 validation sets. We perform a statistical significance test using the $t$-test \cite{cohen2013statistical} to determine whether or not the differences among the 5 random 10-fold cross validations are statistically significant. Recall that differences are considered statistically significant if the significance level (i.e., the $p$-value) is less than or equal to 0.05 (the 95\% confidence level). 
Consider the overall effectiveness F1-measure as an example, Table \ref{Table_t_test} shows the means and standard deviations of the F1-measures for 5 random 10-fold cross-validations. Using the first 10-fold cross-validation as the baseline, the $p$-value column shows the results of the statistical significance test. We observe that all $p$-values are greater than 0.05, meaning that the differences among the 5 random 10-fold cross-validations are not considered statistically significant. Therefore, we use one stratified 10-fold cross-validation in the subsequent experiments.

\begin{table}[!htbp]
\vspace{-0.2cm}
\caption{F1-measures of VulDeeLocator-BGRU for different training sets}
\vspace{-0.2cm}
\label{Table_t_test}
\footnotesize
\centering
\begin{tabular}{|c|c|c|c|}
\hline
\#training sets & Mean  & \tabincell{c}{Standard deviation} & $p$-value\\
\hline
1 & 91.50 & 1.00 & -\\
\hline
2 & 90.59 & 1.25 & 0.09\\
\hline
3 & 90.89 & 0.82 & 0.15\\
\hline
4 & 90.55 & 1.21 & 0.07\\
\hline
5 & 91.93 & 0.88 & 0.32\\
\hline
\end{tabular}
\vspace{-0.2cm}
\end{table}

\begin{insight}
VulDeeLocator leveraging intermediate code-based representation is substantially more effective than VulDeeLocator using source code-based representation, owing to the aforementioned two advantages of intermediate code-based representation.
\end{insight}

\subsection{Experiments for Answering RQ2}
\label{subsec:Experiments_RQ2}
In order to see the capability of BRNN-vdl in locating vulnerabilities, we conduct experiments to compare BRNN-vdl and BRNN while using two types of vulnerability candidates (i.e., source code-based sSeVCs vs. intermediate code-based iSeVCs as specified in Section \ref{subsec:Experiments_RQ1}) and the target programs in Test-set-1 for test.
In what follows, we report the experimental results of using BGRU to instantiate BRNN,
while noting that similar results are observed when using BLSTM to instantiate BRNN.

\begin{table}[!htbp]
\vspace{-0.1cm}
\caption{Comparing BRNN-vdl with BRNN (more specifically, BGRU-vdl vs. BGRU),
where IoU is averaged over the IoUs measured between the detected vulnerable code and the ground-truth vulnerable code in the test data and $|{\sf V}|$ is the average number of detected vulnerable lines of source code.}
\vspace{-0.2cm}
\label{Table_BRNN_vdl_RQ2}
\scriptsize
\centering
\begin{tabular}{|c|c|c|c|c|c|c|c|c|}
\hline
{\tabincell{c}{Vulnerability\\ candidate}} & Model  & {\tabincell{c}{FPR \\(\%)}} & {\tabincell{c}{FNR \\(\%)}} & {\tabincell{c}{A \\(\%)}} & {\tabincell{c}{P \\(\%)}} & {\tabincell{c}{F1 \\(\%)}} & {\tabincell{c}{IoU \\(\%)}} & $|{\sf V}|$\\
\hline
 {\multirow{2}{*}{sSeVC}} & \tabincell{c}{BRNN\\-vdl}  & 5.1 & 11.7 & 92.2 & 92.3 & 90.2 &  25.1 & 2.5\\
 \cline{2-9}
 & BRNN & 10.1 & 12.2 & 89.0 & 84.2 & 86.0 & 8.4 & 17.7\\
\hline
 {\multirow{2}{*}{iSeVC}} & \tabincell{c}{BRNN\\-vdl} & 0.5 & 3.8 & 98.8 & 98.2 & 97.2 & 36.3 & 2.6 \\
\cline{2-9}
 & BRNN & 2.3 & 5.4 & 97.0 & 92.0 & 93.3 & 10.1 & 19.9\\
\hline
\end{tabular}
\vspace{-0.2cm}
\end{table}

Table \ref{Table_BRNN_vdl_RQ2} presents the comparison. 
(i) The BRNN model with sSeVC is the vulnerability detector known as SySeVR \cite{SySeVR}; (ii) the BRNN-vdl model with sSeVC is a variant of our proposed VulDeeLocator by adapting it to deal with source code; (iii) the BRNN model with iSeVC is a variant of SySeVR by adapting it to deal with intermediate code; (iv) the BRNN-vdl model with iSeVC is our proposed VulDeeLocator.
For locating vulnerabilities, BRNN-vdl achieves, on average, a 21.5\% higher IoU than BRNN
because the number of detected vulnerable lines of code is 2.6 for BRNN-vdl and 18.8 for BRNN on average.
This can be explained by the fact that BRNN preserves the input granularity in its output, while BRNN-vdl reduces the input lines of code to much smaller lines of code in its output.
This ``granularity refinement'' is accomplished by the ``vdl'' part.
In terms of vulnerability detection capability,
BRNN-vdl is better than BRNN,
with a 5.0\% lower false-positive rate, a 0.5\% higher false-negative rate, and a 4.2\% higher F1-measure when using sSeVCs as vulnerability candidates, and
with a 1.8\% lower false-positive rate, a 1.6\% false-negative rate, and a 3.9\% higher F1-measure when using iSeVCs as vulnerability candidates.
This means that ``vdl''
can somewhat improve the vulnerability detection capability.
This leads to:

\begin{insight}
BRNN-vdl achieves a substantially higher vulnerability locating precision and a somewhat higher vulnerability detection capability than BRNN.
\end{insight}

\subsection{Experiments for Answering RQ3}
\label{subsec:Experiments_RQ3}

In order to evaluate the effectiveness and preciseness of VulDeeLocator in detecting vulnerabilities with known ground truth,
we compare two instances of VulDeeLocator
and some state-of-the-art vulnerability detectors in terms of their capabilities in detecting and locating vulnerabilities in target programs with known ground truth.

We use the target programs in Test-set-1 and Test-set-2 for test.
Table \ref{Table_Comparison_with_other_tools_detection} summarizes the experimental results.
We observe that on average, VulDeeLocator-BGRU simultaneously achieves a 4.0\% lower false-negative rate, a 2.3\% higher F1-measure, and an 8.3\% higher IoU,
than VulDeeLocator-BLSTM. 
The higher locating precision may be attributed to the fact that BGRU uses fewer parameters, possibly making it easier
to ``refine'' the output. 
We observe that the false-negative rates are higher than the false-positive rates, where the higher false-negative rates may be caused by the inadequate coverage of vulnerability syntax characteristics and vulnerability types in the training data. 
When compared with Test-set-1, Test-set-2 makes VulDeeLocator achieve an 11.5\% higher false-positive rate, a 20.2\% higher false-negative rate, an 18.3\% lower F1-measure, and a 2.8\% lower IoU on average. This is caused by the fact that there are more different features between the training set and Test-set-2. 
For source code- and rule-based vulnerability detectors, we consider the open source tool Flawfinder \cite{FlawFinder} and the commercial product Checkmarx \cite{Checkmarx}.
For intermediate code- and rule-based vulnerability detectors, we consider the commercial product Fortify \cite{HP_Fortify}.
For binary code- and rule-based vulnerability detectors, we consider the open source taint-style vulnerability detector Saluki \cite{gotovchits2018saluki}.
For deep learning-based vulnerability detectors, we consider VulDeePecker \cite{vuldeepecker}, which is designed to detect vulnerabilities related to library/API function calls,
and SySeVR \cite{SySeVR}, which is designed to detect multiple types of vulnerabilities.
The implementations of these two tools are obtained from their authors (via private communications).
We choose these systems for comparison because they are the state-of-the-art and/or available to us.

	\begin{table*}[!htbp]
		
		\caption{Effectiveness of VulDeeLocator-BLSTM, VulDeeLocator-BGRU,
			and state-of-the-art vulnerability detectors, where IoU is averaged over the IoUs measured between the detected vulnerable code and the ground-truth vulnerable code in the test data and $|{\sf V}|$ is the average number of detected vulnerable lines of source code.
		}
		\vspace{-0.2cm}
		\label{Table_Comparison_with_other_tools_detection}
		\footnotesize
		\centering
		\begin{tabular}{|c|c|c|c|c|c|c|c|c|}
			\hline
			Method & Test set & {\tabincell{c}{FPR (\%)}} & {\tabincell{c}{FNR (\%)}} & {\tabincell{c}{A (\%)}} & {\tabincell{c}{P (\%)}} & {\tabincell{c}{F1 (\%)}} & {\tabincell{c}{IoU (\%)}} & $|{\sf V}|$\\
			\hline
			\multicolumn{9}{|c|}{VulDeeLocator with two instances of BRNN} \\
			\hline
			{\multirow{2}{*}{\tabincell{c}{VulDeeLocator\\-BLSTM}}} & Test-set-1 & 0.5  & 7.8 & 97.7 & 98.5 & 95.2 & 27.1 & 2.1\\
			\cline{2-9}
			& Test-set-2 & 12.0 & 28.0 & 81.0 & 81.8 & 76.6 & 25.2 & 3.4\\
			\hline
			{\multirow{2}{*}{\tabincell{c}{VulDeeLocator\\-BGRU}}} & Test-set-1 & 0.5 & 3.8 & 98.8 & 98.2 & 97.2 & 36.3 & 2.6 \\
			\cline{2-9}
			& Test-set-2 & 12.0 & 24.0 & 82.8 & 82.6 & 79.2 & 32.6 & 3.5 \\
			\hline
			\hline
			\multicolumn{9}{|c|}{State-of-the-art vulnerability detectors} \\
			\hline
			{\multirow{2}{*}{Flawfinder}} & Test-set-1 & 10.5 & 83.7 & 61.6 & 48.9 & 24.5 & 38.4 & 5.0\\
			\cline{2-9}
			& Test-set-2 & 24.2 & 78.2 & 53.6 & 38.5 & 27.8 & 35.7 & 5.4\\
			\hline
			{\multirow{2}{*}{Checkmarx}} & Test-set-1 & 72.8 & 54.4 & 39.1 & 27.9 & 34.6 & 26.5 & 4.1\\
			\cline{2-9}
			& Test-set-2 & 66.7 & 56.5 & 37.5 & 31.2 & 36.4 & 18.1 & 6.3\\
			\hline
			{\multirow{2}{*}{Fortify}} & Test-set-1 & 37.0 & 54.0 & 56.5 & 43.4 & 44.6 & 30.9 & 2.4\\
			\cline{2-9}
			& Test-set-2 & 42.4 & 47.8 & 55.4 & 46.2 & 49.0 & 30.5 & 2.9\\
			\hline
			{\multirow{2}{*}{Saluki}} & Test-set-1 & 10.3 & 31.3 & 85.1 & 64.9 & 66.8 & 23.7 & 2.4\\
			\cline{2-9}
			& Test-set-2 & 18.2 & 39.1 & 73.2 & 70.0 & 65.1 & 24.0 & 2.6\\
			\hline
			{\multirow{2}{*}{VulDeePecker}} & Test-set-1 & 7.9 & 49.4 & 51.0 & 91.9 & 65.2 & 9.0 & 14.5\\
			\cline{2-9}
			& Test-set-2 & 21.2 & 48.0 & 67.2 & 65.0 & 57.8 & 8.5 & 98.8\\
			\hline
			{\multirow{2}{*}{SySeVR}} & Test-set-1 & 10.1 & 12.2 & 89.0 & 84.2 & 86.0 & 8.4 & 17.7\\
			\cline{2-9}
			& Test-set-2 & 18.2 & 32.0 & 75.9 & 73.9 & 70.8 & 8.0 & 95.4 \\
			\hline
		\end{tabular}
		\vspace{-0.2cm}
	\end{table*}

Table \ref{Table_Comparison_with_other_tools_detection} summarizes the comparison with state-of-the-art pattern-based vulnerability detectors. We make the following observations.
{\bf (i)} Source code- and rule-based vulnerability detector Flawfinder incurs prohibitively high
false-negative rate, which can be attributed to the inadequacy of its parser and patterns \cite{DBLP:phd/dnb/Yamaguchi15}.
{\bf (ii)} Source code- and rule-based vulnerability detector Checkmarx incurs prohibitively high false-positive rate and false-negative rate despite its use of data-flow analysis; this ineffectiveness can be attributed to the inadequacy of human-written rules. This justifies why we only use the Checkmarx rules to extract sSyVCs as a starting point for vulnerability detection.
{\bf (iii)} Intermediate code- and rule-based vulnerability detector Fortify, which uses a data-flow analysis, incurs very high false-positive rate and false-negative rate, but is better than Checkmarx, suggesting that rules based on intermediate code can indeed accommodate more useful information than rules based on source code.
{\bf (iv)} Binary code- and rule-based vulnerability detector Saluki is more effective than the other rule-based vulnerability detectors because it uses a taint analysis, but is still less effective than deep learning-based detectors. 
Its false negatives and false positives are caused by the incomplete vulnerability rules it used.
{\bf (v)} Deep learning-based detector VulDeePecker is much less effective than deep learning-based SySeVR because the former can only cope with the class of vulnerabilities related to library/API function calls \cite{vuldeepecker}, but SySeVR can cope with multiple classes of vulnerabilities \cite{SySeVR}.
{\bf (vi)} VulDeeLocator-BGRU achieves respectively a 9.8\%, 7.9\%, and 8.2\% improvement over SySeVR in F1-measure, false-positive rate, and false-negative rate on average, 
because it can accommodate more semantic information conveyed by intermediate code.
{\bf (vii)} Rule-based vulnerability detectors (i.e., Flawfinder, Checkmarx, Fortify, and Saluki)
achieve an IoU of 28.5\% on average,
but their low overall effectiveness (F1-measure) hinders their usefulness.
{\bf (viii)} The IoUs of VulDeeLocator-BLSTM and VulDeeLocator-BGRU are much higher than that of VulDeePecker and SySeVR, because the average number of detected vulnerable lines of code is 2.8 for VulDeeLocator-BLSTM and 3.1 for VulDeeLocator-BGRU, while noting that their counterparts are 56.7 for VulDeePecker and 56.6 for SySeVR.
The high vulnerability locating precision can be attributed to the vdl-part of BRNN-vdl.
{\bf (ix)} 
The effectiveness of VulDeePecker and SySeVR when applied to Test-set-2 are much lower than their counterparts when applied to 
Test-set-1, because the training set has more synthetic and academic programs than real-world programs.

\begin{figure}[!htb]
    \centering
    \subfigure[Scores for VulDeeLocator]{
    \label{Fig_VulDeeLocator}
    \includegraphics[width=.32\textwidth]{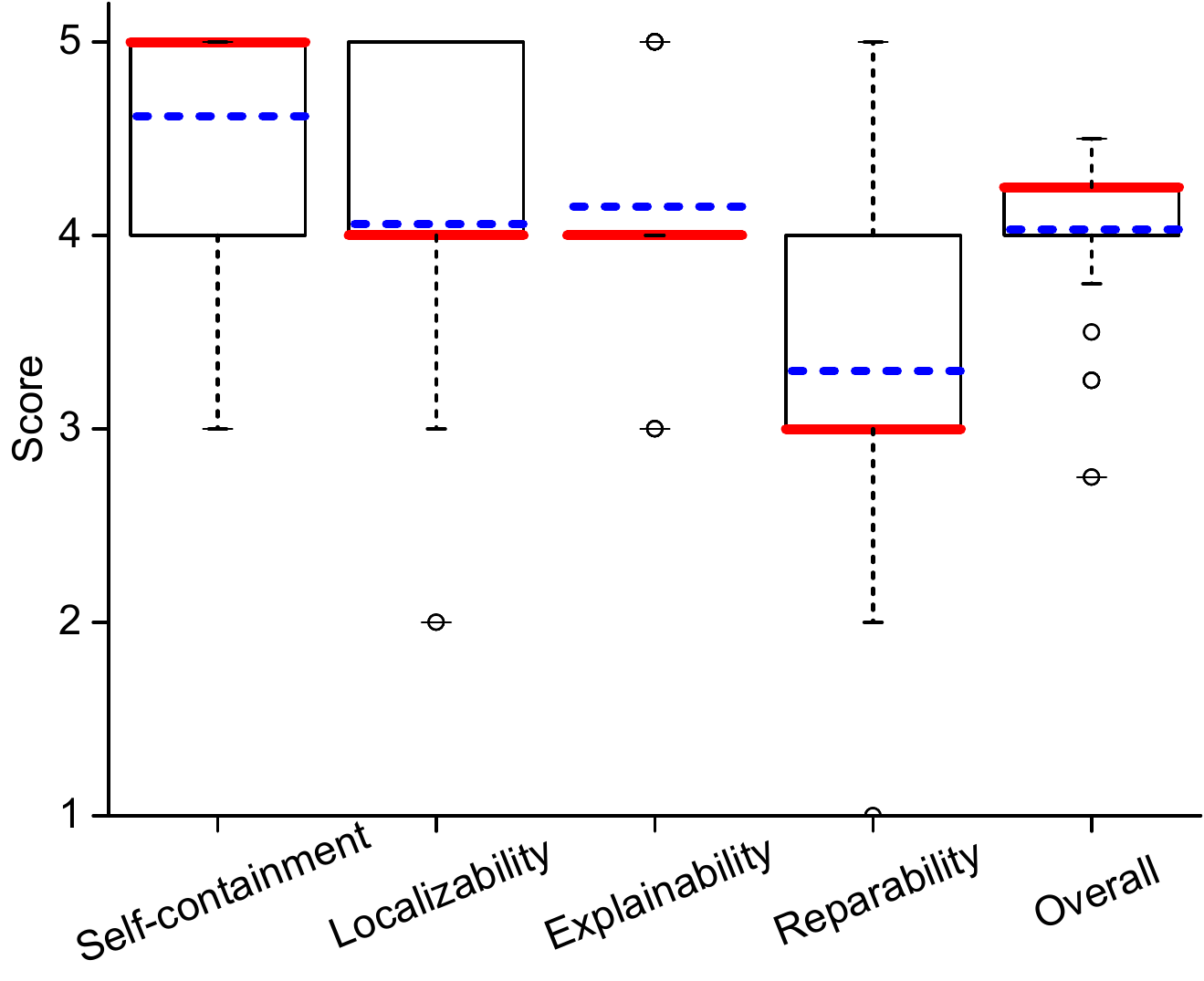}}
     \vspace{-0.2cm}
    \subfigure[Scores for SySeVR]{
    \label{Fig_SySeVR}
    \includegraphics[width=.32\textwidth]{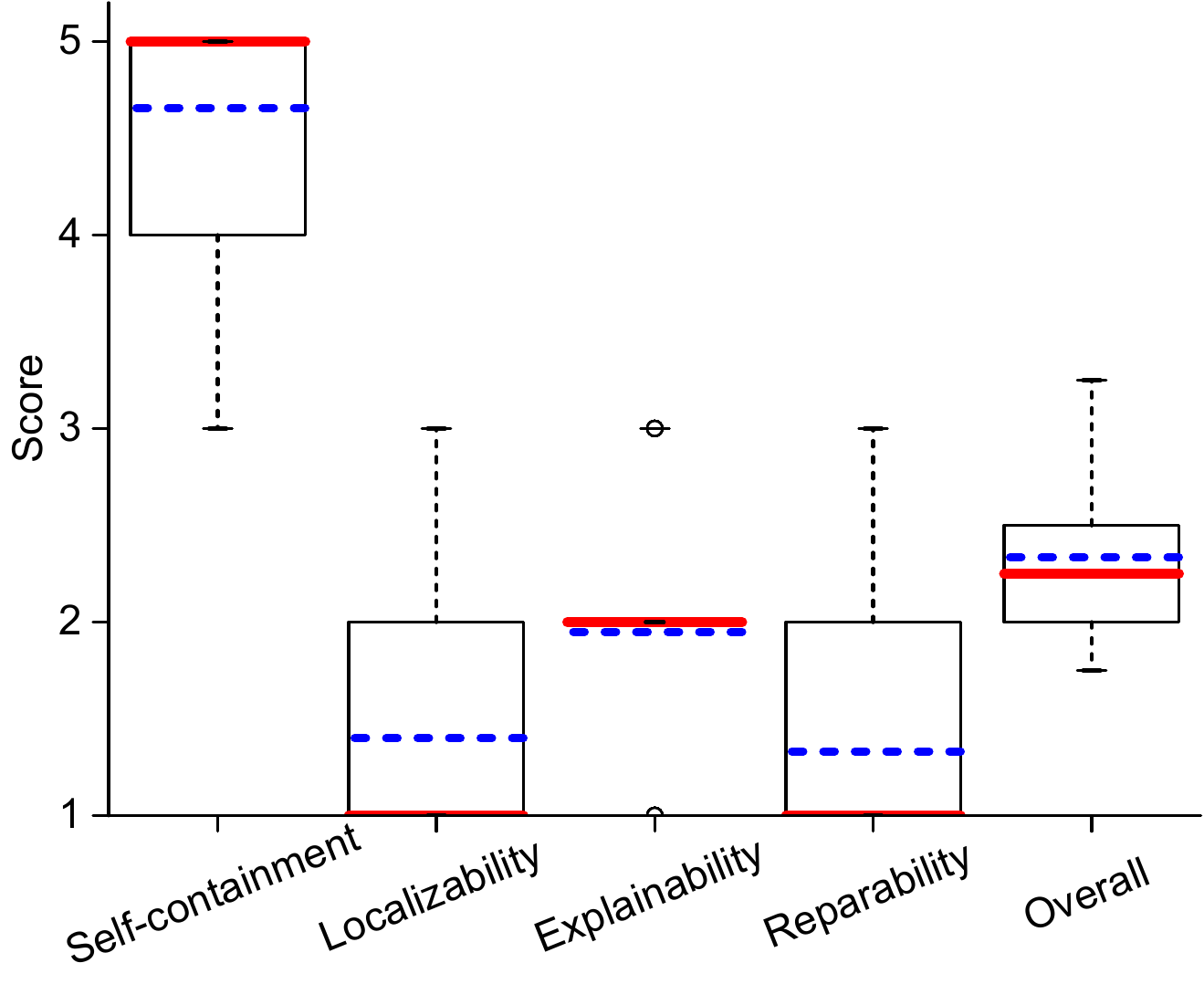}}
     \vspace{-0.1cm}
\caption{Boxplots of the attribute score and the overall score for VulDeeLocator vs. SySeVR, where the solid red line indicates the median and the dashed blue line indicates the mean.} 
    \label{Fig_user_study} 
\end{figure}

\smallskip
\noindent{\bf User study.}
In order to help assess the usability of VulDeeLocator, we perform a user study on VulDeeLocator vs. SySeVR \cite{SySeVR} (i.e., the best state-of-the-art vulnerability detector in Table \ref{Table_Comparison_with_other_tools_detection}), according to the following 4 attributes.
\begin{itemize}
\setlength{\itemsep}{0pt}
\setlength{\parsep}{0pt}
\setlength{\parskip}{2pt}
    \item {\em Self-containment}: 
    This means that the output code fragment of the vulnerability detector is self-contained, meaning that the code fragment itself is sufficient for a user to understand the vulnerability.
    \item{\em Localizability}: The vulnerable lines of code can be identified from the output of the vulnerability detector. 
    \item {\em Explainability}: 
    This is the extent at which a user can explain why the detected vulnerability is indeed a vulnerability.
    \item {\em Reparability}: This is the extent at which a user can come up with a patch to the detected vulnerability.
\end{itemize}
The scale of each attribute score is defined as \{1, 2, 3, 4, 5\}. The overall score of an output code fragment is the average of the 4 attribute scores. The higher the overall score, the better the vulnerability detector. We randomly select 20 vulnerabilities that are correctly detected by VulDeeLocator and 20 vulnerabilities that are correctly detected by SySeVR (i.e., they are all true-positives); and we ask 6 computer science students,  including 3 senior undergraduate and 3 graduate students, to score each of them according to the aforementioned 4 attributes. Fig. \ref{Fig_user_study} presents the boxplots of the 4 attributes score and the overall score.
We observe that the median scores of most attributes for VulDeeLocator are higher ($\geq 2.0$) than those for SySeVR, except for self-containment (which is similar in both cases).
We also observe that the mean scores of most attributes for VulDeeLocator are higher than those for SySeVR, except for self-containment (which is similar in both cases).
The mean overall score for VulDeeLocator (i.e., 4.1) is much higher than that for SySeVR (i.e., 2.3), indicating that VulDeeLocator is better in helping users locate, understand, and fix vulnerabilities. This can be explained by the fact that VulDeeLocator can produce a more focused 
set of code that contains a vulnerability.
In summary, we draw:

\begin{insight}
VulDeeLocator is more effective than the state-of-the-art pattern-based vulnerability detectors in detecting and locating vulnerabilities. In particular, VulDeeLocator-BGRU achieves a 4.2X
higher locating precision than 
vulnerability detector SySeVR on average.
\end{insight}

\subsection{Experiments for Answering RQ4}
\label{subsec:Experiments_RQ4}
In order to show the effectiveness and preciseness of VulDeeLocator in detecting and locating vulnerabilities in real-world software products,
we apply VulDeeLocator-BGRU, which is the most effective instance of VulDeeLocator,
to detect vulnerabilities reported
in the NVD between 2017 and 2019, in 3 real-world software products (i.e., FFmpeg 2.8.2, Wireshark 2.0.5, and Libav 9.10), which are unknown vulnerabilities with respect to the vulnerabilities contained in the training set.
Since we need to manually identify false positives and false negatives for vulnerability detectors, 
which is a time- and labor-consuming process, 
we randomly select 200 (out of the 14,299) 
program files from the main modules of the aforementioned 3 software products, and apply the 5 representative vulnerability detectors (with higher effectiveness as shown in Section \ref{subsec:Experiments_RQ3}) to those program files. 
We collect all of the vulnerabilities detected by the 5 detectors from the 200 program files 
and use them as the ``overall set''. We then  manually examine them to 
identify false positives and further remove the vulnerabilities reported in the NVD prior to 2017.
The experimental results are presented in Table \ref{Table_Comparison_with_other_tools_real_world_software}.  Among the 5 detectors, VulDeeLocator-BGRU is the most effective. Specifically, VulDeeLocator-BGRU detects 22 vulnerabilities from the 200 program files, including 18 confirmed vulnerabilities (i.e., trust positives) and 4 false positives, while missing 5 vulnerabilities. 
Among the other 4 detectors, SySeVR achieves the highest F1 but the lowest IoU. The low precise location of SySeVR is caused by the large average number of detected vulnerable lines of code (i.e., on average, 76.2 lines of code per detected vulnerability), which is in sharp contrast to VulDeeLocator’s average number of detected vulnerable lines of code (i.e., on average, 3.8 lines of code per detected vulnerability).

Table \ref{Table_known_vulnerabilities} describes the vulnerabilities that are confirmed or missed by VulDeeLocator-BGRU. Among 18 confirmed one, 2 belong to Libav 9.10 and are not reported in the NVD but have been ``silently'' patched by the vendor when releasing newer versions. 
For example, the vulnerability in matroska\-dec.c is a use-after-free vulnerability related to  pointer ``$tracks$;'' this vulnerability is not reported in the NVD but has been ``silently'' patched by the vendor when releasing Libav 9.18 and later versions.

\begin{insight}
VulDeeLocator can detect and pinpoint vulnerabilities in real-world software products.
\end{insight}

\begin{table}[!htbp]
		\caption{Effectiveness of vulnerability detectors when applied to the 200 program files that are selected from the 3 real-world software products, where $|{\sf V}|$ is the average number of detected vulnerable lines of code.
		}
		\vspace{-0.2cm}
		\label{Table_Comparison_with_other_tools_real_world_software}
		\footnotesize
		\centering
		\begin{tabular}{|c|c|c|c|c|c|c|c|}
			\hline
			Method & {\tabincell{c}{FPR \\(\%)}} & {\tabincell{c}{FNR \\(\%)}} & {\tabincell{c}{A \\(\%)}} & {\tabincell{c}{P \\(\%)}} & {\tabincell{c}{F1 \\(\%)}} & {\tabincell{c}{IoU \\(\%)}} & $|{\sf V}|$\\
			\hline
			\tabincell{c}{VulDeeLocator\\-BGRU} & 12.1 & 21.7 & 83.9 & 81.8 & 80.0 & 30.2 & 3.8 \\
			\hline
			Checkmarx & 60.6 & 56.5 & 41.1 & 33.3 & 37.7 & 24.2 & 6.0 \\
			\hline
			Fortify & 45.5 & 47.8 & 53.6 & 44.4 & 48.0 & 28.1 & 2.7 \\
			\hline
			Saluki & 21.2 & 39.1 & 71.4 & 66.7 & 63.6 & 25.2 & 3.2 \\
			\hline
			SySeVR & 18.1 & 34.8 & 77.0 & 71.4 & 68.2 & 8.5 & 76.2 \\
			\hline
		\end{tabular}
		\vspace{-0.2cm}
	\end{table}

\begin{table}[!htbp]
\vspace{-0.2cm}
\caption{The 18 vulnerabilities that are correctly detected and the 5 vulnerabilities that are missed by VulDeeLocator-BGRU (from the 200 program files)} 
\vspace{-0.2cm}
\label{Table_known_vulnerabilities}
\centering
\scriptsize
\begin{tabular}{|c|c|c|c|}
\hline
\tabincell{c}{Target\\ product} & CVE ID & \tabincell{c}{Vulnerable file } & \tabincell{c}{Status}  \\
\hline
{\multirow{6}{*}{\tabincell{c}{FFmpeg\\ 2.8.2}}} & CVE-2017-9608 & .../dnxhd\_parser.c & Confirmed \\
\cline{2-4}
 & CVE-2018-14394 & .../movenc.c & Confirmed  \\
\cline{2-4}
 & CVE-2018-14395 & .../movenc.c & Confirmed \\
 \cline{2-4}
 & CVE-2018-1999010 & .../mms.c & Confirmed \\
 \cline{2-4}
 & CVE-2019-12730 & .../aadec.c & Confirmed \\
 \cline{2-4}
 & CVE-2017-9996 & .../cdxl.c & Missed \\
 \hline
{\multirow{11}{*}{\tabincell{c}{Wireshark\\ 2.0.5}}} & CVE-2017-6467 & .../rmap.c & Confirmed \\
\cline{2-4}
& CVE-2017-6468 & .../netscaler.c & Confirmed \\
\cline{2-4}
& CVE-2017-6470 & .../packet-iax2.c & Confirmed \\
\cline{2-4}
& CVE-2017-6474 & .../netscaler.c & Confirmed \\
\cline{2-4}
& CVE-2017-7700 & .../netscaler.c & Confirmed \\
\cline{2-4}
& CVE-2017-9345 & .../packet-dns.c & Confirmed \\
\cline{2-4}
& CVE-2017-11410 & .../packet-wbxml.c & Confirmed \\
\cline{2-4}
& CVE-2017-11411 & .../packet-opensafety.c & Confirmed \\
\cline{2-4}
& CVE-2017-13767 & .../packet-msdp.c & Confirmed \\
\cline{2-4}
& CVE-2017-9344 & .../packet-btl2cap.c & Missed \\
\cline{2-4}
& CVE-2017-13766 & .../packet-dcerpc-pn-io.c & Missed \\
\hline
{\multirow{6}{*}{\tabincell{c}{Libav 9.10}}} & - & .../matroska-dec.c & Confirmed \\
\cline{2-4}
& - & .../pngdsp.c & Confirmed \\
\cline{2-4}
& CVE-2018-5766 & .../avpacket.c & Confirmed \\
\cline{2-4}
& CVE-2018-5684 & .../mov.c & Confirmed \\
\cline{2-4}
& CVE-2017-16803 & .../smacker.c & Missed \\
\cline{2-4}
& CVE-2017-9051 & .../nsvdec.c & Missed \\
\hline
 \end{tabular}
 \vspace{-0.2cm}
\end{table}

%% file: limitations.tex
This study has several limitations.
{\bf First}, the design of VulDeeLocator focuses on detecting vulnerabilities in C source programs
because (i) we want to demonstrate the feasibility of VulDeeLocator and (ii) the
tools we leverage happen to support C. Extending VulDeeLocator to accommodate other programming languages is an interesting future work.
{\bf Second}, VulDeeLocator requires to compile program source code into intermediate code, and cannot be used when a program source code cannot be compiled.
{\bf Third}, the four kinds of vulnerability syntax characteristics used by VulDeeLocator can cover 98.3\% of vulnerable programs collected from NVD
and SARD. This 98.3\% coverage should be used with caution
because (i) for the NVD data,
we only use the lines of code that are deleted or moved in a diff file as the location of a vulnerability (i.e., we did not consider those vulnerabilities whose diff files only involve line additions),
and (ii) the SARD data may not be representative of real-world software products.
It is an open problem to identify more complete vulnerability syntax characteristics.
{\bf Fourth}, our case study uses BRNN-vdl to instantiate VulDeeLocator to demonstrate  feasibility. Tailored neural networks need to be designed for vulnerability detection purposes.
{\bf Fifth}, VulDeeLocator, as a static vulnerability detector, cannot accurately detect vulnerabilities that depend on dynamic information during program running.
{\bf Sixth}, we can partly explain the effectiveness of VulDeeLocator, but much more research needs to be done in this direction of {\em explainability}.

%% file: related_work.tex
\noindent{\bf Prior work on static vulnerability detection.}
The present study belongs to static vulnerability detection, which includes {\em code simila\-rity-based} methods and {\em pattern-based} methods.
Code similarity-based methods
\cite{kim2017vuddy,jang2012redebug,li2016vulpecker} 
can achieve a high locating precision when they indeed detect vulnerabilities,
but have a high false-negative rate because many vulnerabilities are not caused by code cloning \cite{vuldeepecker}.
Pattern-based vulnerability detection methods
can be further divided into {\em rule-based} ones and {\em machine learning-based} ones.
Rule-based methods use analyst-generated rules to detect vulnerabilities, including (i) open source tools (e.g., Flawfinder \cite{FlawFinder}) and commercial tools (e.g., Checkmarx \cite{Checkmarx}), which operate on program source code, and (ii) Fortify and Coverity \cite{HP_Fortify,Coverity}, which operate on intermediate code. These tools
have high false-positives or false-negatives \cite{DBLP:phd/dnb/Yamaguchi15}.
Machine learning-based methods aim to detect vulnerabilities using patterns learned from analyst-defined feature representations of vulnerabilities
\cite{yamaguchi2012generalized,neuhaus2007predicting,grieco2016toward,yamaguchi2013chucky,yamaguchi2015automatic,DBLP:conf/icse/Du0LGZLJ19}
or ``raw'' feature representations via deep learning \cite{DBLP:conf/ccs/LinZLPX17,DBLP:journals/corr/abs-1807-04320,vuldeepecker,SySeVR,DBLP:journals/tii/LinZLPXVM18,duan2019vulsniper,zhou2019devign,DBLP:journals/corr/abs-2001-02334}. 
These methods detect vulnerabilities at coarse granularities (e.g., program \cite{grieco2016toward}, component \cite{neuhaus2007predicting}, function
\cite{yamaguchi2012generalized,yamaguchi2013chucky,DBLP:journals/corr/abs-1807-04320,DBLP:conf/ccs/LinZLPX17,DBLP:journals/tii/LinZLPXVM18,duan2019vulsniper,zhou2019devign}, and program slice
\cite{vuldeepecker,SySeVR,DBLP:journals/corr/abs-2001-02334}). In addition, there are some loosely related work, such as DeepSim \cite{DBLP:conf/sigsoft/ZhaoH18} in the sense of using intermediate representation but for Java code similarity and Instruction2vec \cite{lee2019instruction2vec} in the sense of detecting vulnerabilities in binary code by modeling assembly code. 

Among the detectors mentioned above, VulDeePecker \cite{vuldeepecker}, SySeVR \cite{SySeVR}, and $\mu$VulDeePecker \cite{DBLP:journals/corr/abs-2001-02334} are closely related to ours because they are also based on program slices.
However, these detectors have little capability in accommodating semantic information (e.g., relations between the definitions of types and macros and their uses across files, and control flows and variable define-use relations).
Moreover, they cannot precisely pin down the locations of vulnerabilities. Nevertheless, $\mu$VulDeePecker \cite{DBLP:journals/corr/abs-2001-02334} aims to detect specific types of vulnerabilities in source code.
VulDeeLocater moves a significant step forward by tackling the newly articulated requirements of achieving high detection capability and high locating precision in vulnerability detection.

\noindent{\bf Prior work on dynamic vulnerability detection.}
Dynamic vulnerability detection, including dynamic symbolic execution \cite{DBLP:conf/kbse/AlatawiSM17} 
and fuzzing \cite{DBLP:conf/sp/GanZQTLPC18}, 
is complementary to static vulnerability detection and is often used to detect vulnerabilities in binary code. These methods explore program execution paths to identify the inputs that make the program exhibit unsafe operations (e.g., crashing).
These methods leverage some finite sets of execution traces to determine whether or not a program is buggy or vulnerable, meaning that they may miss many vulnerabilities.
In contrast, VulDeeLocator can scan various
paths in the program and can possibly detect more vulnerabilities. 

\smallskip

\noindent{\bf Prior work on bug detection.}
Since vulnerabilities can be seen as a special kind of bugs \cite{DBLP:journals/chinaf/SunPZLC19}, we briefly review prior studies on bug detectors.
Similar to vulnerability detection, there are two detection methods: static vs. dynamic.
Static methods often use information retrieval techniques together with bug reports to detect bugs in source code (e.g., \cite{DBLP:conf/ease/XiaoKMB18}).
Dynamic methods include:
spectrum-based methods \cite{DBLP:journals/tosem/XieCKX13}, which examine pass-and-fail execution traces to determine whether or not a line of source code has a bug;
mutation-based methods \cite{DBLP:journals/stvr/PapadakisT15},
which consider whether or not the execution of a line of code affects the result of a test case.
However, bug detection methods cannot be used to detect vulnerabilities because (i) bugs are not necessarily vulnerabilities and (ii) bug detection methods often rely on bug reports or test cases.

%% file: bare_jrnl_compsoc.bbl
\begin{thebibliography}{10}
\providecommand{\url}[1]{#1}
\csname url@samestyle\endcsname
\providecommand{\newblock}{\relax}
\providecommand{\bibinfo}[2]{#2}
\providecommand{\BIBentrySTDinterwordspacing}{\spaceskip=0pt\relax}
\providecommand{\BIBentryALTinterwordstretchfactor}{4}
\providecommand{\BIBentryALTinterwordspacing}{\spaceskip=\fontdimen2\font plus
\BIBentryALTinterwordstretchfactor\fontdimen3\font minus
  \fontdimen4\font\relax}
\providecommand{\BIBforeignlanguage}[2]{{%
\expandafter\ifx\csname l@#1\endcsname\relax
\typeout{** WARNING: IEEEtran.bst: No hyphenation pattern has been}%
\typeout{** loaded for the language `#1'. Using the pattern for}%
\typeout{** the default language instead.}%
\else
\language=\csname l@#1\endcsname
\fi
#2}}
\providecommand{\BIBdecl}{\relax}
\BIBdecl

\bibitem{CVE}
\emph{{CVE}}, \url{http://cve.mitre.org/}.

\bibitem{kim2017vuddy}
S.~Kim, S.~Woo, H.~Lee, and H.~Oh, ``{VUDDY}: A scalable approach for
  vulnerable code clone discovery,'' in \emph{Proceedings of 2017 {IEEE}
  Symposium on Security and Privacy, San Jose, CA, USA}, 2017, pp. 595--614.

\bibitem{jang2012redebug}
J.~Jang, A.~Agrawal, and D.~Brumley, ``{ReDeBug}: Finding unpatched code clones
  in entire {OS} distributions,'' in \emph{Proceedings of 2012 {IEEE} Symposium
  on Security and Privacy, San Francisco, California, {USA}}, 2012, pp. 48--62.

\bibitem{li2016vulpecker}
Z.~Li, D.~Zou, S.~Xu, H.~Jin, H.~Qi, and J.~Hu, ``{VulPecker}: An automated
  vulnerability detection system based on code similarity analysis,'' in
  \emph{Proceedings of the 32nd Annual Conference on Computer Security
  Applications, Los Angeles, CA, USA}, 2016, pp. 201--213.

\bibitem{FlawFinder}
\emph{{Flawfinder}}, \url{https://dwheeler.com/flawfinder/}.

\bibitem{Checkmarx}
\emph{Checkmarx}, \url{https://www.checkmarx.com/}.

\bibitem{HP_Fortify}
\emph{{Fortify}},
  \url{https://www.microfocus.com/en-us/portfolio/application-security}.

\bibitem{Coverity}
\emph{Coverity}, \url{https://scan.coverity.com/}.

\bibitem{DBLP:journals/ijndc/LiangWWX16}
H.~Liang, L.~Wang, D.~Wu, and J.~Xu, ``{MLSA:} a static bugs analysis tool
  based on {LLVM} {IR},'' in \emph{Proceedings of the 17th {IEEE/ACIS}
  International Conference on Software Engineering, Artificial Intelligence,
  Networking and Parallel/Distributed Computing, Shanghai, China}, 2016, pp.
  407--412.

\bibitem{DBLP:journals/chinaf/FangLZWWW17}
Z.~Fang, Q.~Liu, Y.~Zhang, K.~Wang, Z.~Wang, and Q.~Wu, ``A static technique
  for detecting input validation vulnerabilities in {Android} apps,''
  \emph{{SCIENCE} {CHINA} Information Sciences}, vol.~60, no.~5, pp.
  052\,111:1--052\,111:16, 2017.

\bibitem{yamaguchi2012generalized}
F.~Yamaguchi, M.~Lottmann, and K.~Rieck, ``Generalized vulnerability
  extrapolation using abstract syntax trees,'' in \emph{Proceedings of the 28th
  Annual Computer Security Applications Conference, Orlando, FL, USA}, 2012,
  pp. 359--368.

\bibitem{neuhaus2007predicting}
S.~Neuhaus, T.~Zimmermann, C.~Holler, and A.~Zeller, ``Predicting vulnerable
  software components,'' in \emph{Proceedings of 2007 {ACM} Conference on
  Computer and Communications Security, Alexandria, Virginia, USA}, 2007, pp.
  529--540.

\bibitem{grieco2016toward}
G.~Grieco, G.~L. Grinblat, L.~C. Uzal, S.~Rawat, J.~Feist, and L.~Mounier,
  ``Toward large-scale vulnerability discovery using machine learning,'' in
  \emph{Proceedings of the 6th {ACM} on Conference on Data and Application
  Security and Privacy, New Orleans, LA, USA}, 2016, pp. 85--96.

\bibitem{yamaguchi2013chucky}
F.~Yamaguchi, C.~Wressnegger, H.~Gascon, and K.~Rieck, ``Chucky: Exposing
  missing checks in source code for vulnerability discovery,'' in
  \emph{Proceedings of 2013 {ACM} {SIGSAC} Conference on Computer and
  Communications Security, Berlin, Germany}, 2013, pp. 499--510.

\bibitem{yamaguchi2015automatic}
F.~Yamaguchi, A.~Maier, H.~Gascon, and K.~Rieck, ``Automatic inference of
  search patterns for taint-style vulnerabilities,'' in \emph{Proceedings of
  2015 {IEEE} Symposium on Security and Privacy, San Jose, CA, USA}, 2015, pp.
  797--812.

\bibitem{vuldeepecker}
Z.~Li, D.~Zou, S.~Xu, X.~Ou, H.~Jin, S.~Wang, Z.~Deng, and Y.~Zhong,
  ``{VulDeePecker}: A deep learning-based system for vulnerability detection,''
  in \emph{Proceedings of the 25th Annual Network and Distributed System
  Security Symposium, San Diego, California, USA}, 2018.

\bibitem{SySeVR}
Z.~Li, D.~Zou, S.~Xu, H.~Jin, Y.~Zhu, and Z.~Chen, ``{SySeVR}: {A} framework
  for using deep learning to detect software vulnerabilities,'' \emph{{IEEE}
  Trans. Dependable Sec. Comput.}, vol.~PP, pp. 1--1, 2021.

\bibitem{DBLP:journals/corr/abs-2001-02334}
D.~Zou, S.~Wang, S.~Xu, Z.~Li, and H.~Jin, ``{\(\mu\)}{VulDeePecker}: A deep
  learning-based system for multiclass vulnerability detection,'' \emph{{IEEE}
  Trans. Dependable Sec. Comput.}, vol.~PP, pp. 1--1, 2019.

\bibitem{DBLP:conf/ccs/LinZLPX17}
G.~Lin, J.~Zhang, W.~Luo, L.~Pan, and Y.~Xiang, ``{POSTER}: Vulnerability
  discovery with function representation learning from unlabeled projects,'' in
  \emph{Proceedings of 2017 {ACM} {SIGSAC} Conference on Computer and
  Communications Security, Dallas, TX, USA}, 2017, pp. 2539--2541.

\bibitem{DBLP:journals/corr/abs-1807-04320}
R.~L. Russell, L.~Y. Kim, L.~H. Hamilton, T.~Lazovich, J.~A. Harer, O.~Ozdemir,
  P.~M. Ellingwood, and M.~W. McConley, ``Automated vulnerability detection in
  source code using deep representation learning,'' in \emph{Proceedings of the
  17th {IEEE} International Conference on Machine Learning and Applications,
  Orlando, FL, USA}, 2018, pp. 757--762.

\bibitem{DBLP:journals/tii/LinZLPXVM18}
G.~Lin, J.~Zhang, W.~Luo, L.~Pan, Y.~Xiang, O.~Y. de~Vel, and P.~Montague,
  ``Cross-project transfer representation learning for vulnerable function
  discovery,'' \emph{{IEEE} Trans. Industrial Informatics}, vol.~14, no.~7, pp.
  3289--3297, 2018.

\bibitem{duan2019vulsniper}
X.~Duan, J.~Wu, S.~Ji, Z.~Rui, T.~Luo, M.~Yang, and Y.~Wu, ``{VulSniper}: Focus
  your attention to shoot fine-grained vulnerabilities,'' in \emph{Proceedings
  of the 28th International Joint Conference on Artificial Intelligence, Macao,
  China}, 2019, pp. 4665--4671.

\bibitem{zhou2019devign}
Y.~Zhou, S.~Liu, J.~Siow, X.~Du, and Y.~Liu, ``Devign: Effective vulnerability
  identification by learning comprehensive program semantics via graph neural
  networks,'' in \emph{Proceedings of Annual Conference on Neural Information
  Processing Systems}, 2019, pp. 10\,197--10\,207.

\bibitem{SlicingLLVMBitcode}
M.~Chalupa, ``Slicing of {LLVM} bitcode,'' Master's thesis, Masaryk University,
  2016.

\bibitem{DBLP:conf/icse/Du0LGZLJ19}
X.~Du, B.~Chen, Y.~Li, J.~Guo, Y.~Zhou, Y.~Liu, and Y.~Jiang, ``Leopard:
  Identifying vulnerable code for vulnerability assessment through program
  metrics,'' in \emph{Proceedings of the 41st International Conference on
  Software Engineering, Montreal, QC, Canada}, 2019, pp. 60--71.

\bibitem{NVD}
\emph{{NVD}}, \url{https://nvd.nist.gov/}.

\bibitem{DBLP:books/mk/Muchnick1997}
S.~S. Muchnick, \emph{Advanced compiler design and implementation}.\hskip 1em
  plus 0.5em minus 0.4em\relax Morgan Kaufmann, 1997.

\bibitem{DBLP:journals/jpl/Tip95}
F.~Tip, ``A survey of program slicing techniques,'' \emph{Journal of
  Programming Languages}, vol.~3, no.~3, 1995.

\bibitem{DBLP:conf/cgo/LattnerA04}
C.~Lattner and V.~S. Adve, ``{LLVM: A} compilation framework for lifelong
  program analysis {\&} transformation,'' in \emph{Proceedings of the 2nd
  {IEEE}/{ACM} International Symposium on Code Generation and Optimization, San
  Jose, CA, {USA}}, 2004, pp. 75--88.

\bibitem{DG}
\emph{dg}, \url{https://github.com/mchalupa/dg}.

\bibitem{DBLP:journals/csur/PendletonGCX17}
M.~Pendleton, R.~Garcia{-}Lebron, J.~Cho, and S.~Xu, ``A survey on systems
  security metrics,'' \emph{{ACM} Comput. Surv.}, vol.~49, no.~4, pp.
  62:1--62:35, 2017.

\bibitem{IoU}
\emph{{Intersection over Union}},
  \url{https://en.wikipedia.org/wiki/Jaccard_index?tdsourcetag=s_pctim_aiomsg}.

\bibitem{SARD}
\emph{{Software Assurance Reference Dataset}},
  \url{https://samate.nist.gov/SRD/index.php}.

\bibitem{word2vec}
\emph{word2vec}, \url{http://radimrehurek.com/gensim/models/word2vec.html}.

\bibitem{yamaguchi2014modeling}
F.~Yamaguchi, N.~Golde, D.~Arp, and K.~Rieck, ``Modeling and discovering
  vulnerabilities with code property graphs,'' in \emph{Proceedings of 2014
  {IEEE} Symposium on Security and Privacy, Berkeley, USA}, 2014, pp. 590--604.

\bibitem{mani2003knn}
I.~Mani and I.~Zhang, ``{KNN} approach to unbalanced data distributions: A case
  study involving information extraction,'' in \emph{Proceedings of ICML
  Workshop on Learning from Imbalanced Datasets}, 2003, pp. 42--48.

\bibitem{chawla2002smote}
N.~V. Chawla, K.~W. Bowyer, L.~O. Hall, and W.~P. Kegelmeyer, ``{SMOTE}:
  Synthetic minority over-sampling technique,'' \emph{J. Artif. Intell. Res.},
  vol.~16, pp. 321--357, 2002.

\bibitem{cohen2013statistical}
J.~Cohen, \emph{Statistical Power Analysis for the Behavioral Sciences}.\hskip
  1em plus 0.5em minus 0.4em\relax Academic press, 2013.

\bibitem{gotovchits2018saluki}
I.~Gotovchits, R.~Van~Tonder, and D.~Brumley, ``Saluki: Finding taint-style
  vulnerabilities with static property checking,'' in \emph{Proceedings of the
  NDSS Workshop on Binary Analysis Research, San Diego, CA, USA}, 2018.

\bibitem{DBLP:phd/dnb/Yamaguchi15}
F.~Yamaguchi, ``Pattern-based vulnerability discovery,'' Ph.D. dissertation,
  University of G{\"{o}}ttingen, 2015.

\bibitem{DBLP:conf/sigsoft/ZhaoH18}
G.~Zhao and J.~Huang, ``{DeepSim}: Deep learning code functional similarity,''
  in \emph{Proceedings of 2018 {ACM} Joint Meeting on European Software
  Engineering Conference and Symposium on the Foundations of Software
  Engineering, Lake Buena Vista, FL, USA}, 2018, pp. 141--151.

\bibitem{lee2019instruction2vec}
Y.~Lee, H.~Kwon, S.-H. Choi, S.-H. Lim, S.~H. Baek, and K.-W. Park,
  ``Instruction2vec: Efficient preprocessor of assembly code to detect software
  weakness with {CNN},'' \emph{Applied Sciences}, vol.~9, no.~19, p. 4086,
  2019.

\bibitem{DBLP:conf/kbse/AlatawiSM17}
E.~Alatawi, H.~S{\o}ndergaard, and T.~Miller, ``Leveraging abstract
  interpretation for efficient dynamic symbolic execution,'' in
  \emph{Proceedings of the 32nd {IEEE/ACM} International Conference on
  Automated Software Engineering, Urbana, IL, USA}, 2017, pp. 619--624.

\bibitem{DBLP:conf/sp/GanZQTLPC18}
S.~Gan, C.~Zhang, X.~Qin, X.~Tu, K.~Li, Z.~Pei, and Z.~Chen, ``{CollAFL}: Path
  sensitive fuzzing,'' in \emph{Proceedings of 2018 {IEEE} Symposium on
  Security and Privacy, San Francisco, California, {USA}}, 2018, pp. 679--696.

\bibitem{DBLP:journals/chinaf/SunPZLC19}
X.~Sun, X.~Peng, K.~Zhang, Y.~Liu, and Y.~Cai, ``How security bugs are fixed
  and what can be improved: an empirical study with mozilla,'' \emph{{SCIENCE}
  {CHINA} Information Sciences}, vol.~62, no.~1, pp. 19\,102:1--19\,102:3,
  2019.

\bibitem{DBLP:conf/ease/XiaoKMB18}
Y.~Xiao, J.~Keung, Q.~Mi, and K.~E. Bennin, ``Bug localization with semantic
  and structural features using convolutional neural network and cascade
  forest,'' in \emph{Proceedings of the 22nd International Conference on
  Evaluation and Assessment in Software Engineering, Christchurch, New
  Zealand}, 2018, pp. 101--111.

\bibitem{DBLP:journals/tosem/XieCKX13}
X.~Xie, T.~Y. Chen, F.~Kuo, and B.~Xu, ``A theoretical analysis of the risk
  evaluation formulas for spectrum-based fault localization,'' \emph{ACM
  Transactions on Software Engineering and Methodology}, vol.~22, no.~4, pp.
  31:1--31:40, 2013.

\bibitem{DBLP:journals/stvr/PapadakisT15}
M.~Papadakis and Y.~L. Traon, ``{Metallaxis-FL}: mutation-based fault
  localization,'' \emph{Software Testing, Verification \& Reliability},
  vol.~25, no. 5-7, pp. 605--628, 2015.

\end{thebibliography}
